\documentclass[aps,prc,twocolumn,showpacs,superscriptaddress,letterpaper]{revtex4}

\usepackage{graphicx}
\usepackage{dcolumn}
\usepackage{bm}
\usepackage{longtable}
\usepackage{ifthen}
\usepackage{slashbox}

\newcommand{\twocol}{true}

\begin{document}

\title{Electroproduction of $\eta$ Mesons in the $S_{11}(1535)$ Resonance Region at High Momentum Transfer}

\author{M.M.~Dalton}
\email{dalton@jlab.org}
\affiliation{University of the Witwatersrand, Johannesburg, South Africa}
\author{G.S.~Adams}
\affiliation{Rensselaer Polytechnic Institute, Troy, New York 12180}
\author{A.~Ahmidouch}
\affiliation{North Carolina A \& T State University, Greensboro, North Carolina 27411}
\author{T.~Angelescu}
\affiliation{Bucharest University, Bucharest, Romania}
\author{J.~Arrington}
\affiliation{Argonne National Laboratory, Argonne, Illinois 60439}
\author{R.~Asaturyan}
\thanks{Deceased}
\affiliation{Yerevan Physics Institute, Yerevan, Armenia}
\author{O.K.~Baker}
\affiliation{Hampton University, Hampton, Virginia 23668}
\affiliation{Thomas Jefferson National Accelerator Facility, Newport News, Virginia 23606}
\author{N.~Benmouna}
\affiliation{The George Washington University, Washington, D.C. 20052}
\author{C.~Bertoncini}
\affiliation{Vassar College, Poughkeepsie, New York 12604}
\author{W.U.~Boeglin}
\affiliation{Florida International University, University Park, Florida 33199}
\author{P.E.~Bosted}
\affiliation{Thomas Jefferson National Accelerator Facility, Newport News, Virginia 23606}
\author{H.~Breuer}
\affiliation{University of Maryland, College Park, Maryland 20742}
\author{M.E.~Christy}
\affiliation{Hampton University, Hampton, Virginia 23668}
\author{S.H.~Connell}
\affiliation{University of Johannesburg, Johannesburg, South Africa}
\author{Y.~Cui}
\affiliation{University of Houston, Houston, TX 77204}
\author{S.~Danagoulian}
\affiliation{North Carolina A \& T State University, Greensboro, North Carolina 27411}
\author{D.~Day}
\affiliation{University of Virginia, Charlottesville, Virginia 22901}
\author{T.~Dodario}
\affiliation{University of Houston, Houston, TX 77204}
\author{J.A.~Dunne}
\affiliation{Mississippi State University, Mississippi State, Mississippi 39762}
\author{D.~Dutta}
\affiliation{Mississippi State University, Mississippi State, Mississippi 39762}
\author{N.~El~Khayari}
\affiliation{University of Houston, Houston, TX 77204}
\author{R.~Ent}
\affiliation{Thomas Jefferson National Accelerator Facility, Newport News, Virginia 23606}
\author{H.C.~Fenker}
\affiliation{Thomas Jefferson National Accelerator Facility, Newport News, Virginia 23606}
\author{V.V.~Frolov}
\affiliation{LIGO Livingston Observatory, Livingston, LA 70754}
\author{L.~Gan}
\affiliation{University of North Carolina Wilmington, Wilmington, North Carolina 28403}
\author{D.~Gaskell}
\affiliation{Thomas Jefferson National Accelerator Facility, Newport News, Virginia 23606}
\author{K.~Hafidi}
\affiliation{Argonne National Laboratory, Argonne, Illinois 60439}
\author{W.~Hinton}
\affiliation{Hampton University, Hampton, Virginia 23668}
\affiliation{Thomas Jefferson National Accelerator Facility, Newport News, Virginia 23606}
\author{R.J.~Holt}
\affiliation{Argonne National Laboratory, Argonne, Illinois 60439}
\author{T.~Horn}
\affiliation{University of Maryland, College Park, Maryland 20742}
\author{G.M.~Huber}
\affiliation{University of Regina, Regina, Saskatchewan, Canada, S4S 0A2}
\author{E.~Hungerford}
\affiliation{University of Houston, Houston, TX 77204}
\author{X.~Jiang}
\affiliation{Rutgers, The State University of New Jersey, Piscataway, New Jersey, 08855}
\author{M.K.~Jones}
\affiliation{Thomas Jefferson National Accelerator Facility, Newport News, Virginia 23606}
\author{K.~Joo}
\affiliation{University of Connecticut, Storrs, Connecticut 06269}
\author{N.~Kalantarians}
\affiliation{University of Houston, Houston, TX 77204}
\author{J.J.~Kelly}
\thanks{Deceased}
\affiliation{University of Maryland, College Park, Maryland 20742}
\author{C.E.~Keppel}
\affiliation{Hampton University, Hampton, Virginia 23668}
\affiliation{Thomas Jefferson National Accelerator Facility, Newport News, Virginia 23606}
\author{V.~Kubarovsky}
\affiliation{Thomas Jefferson National Accelerator Facility, Newport News, Virginia 23606}
\affiliation{Rensselaer Polytechnic Institute, Troy, New York 12180}
\author{Y.~Li}
\affiliation{Hampton University, Hampton, Virginia 23668}
\author{Y.~Liang}
\affiliation{Ohio University, Athens, Ohio 45071}
\author{S.~Malace}
\affiliation{Bucharest University, Bucharest, Romania}
\author{P.~Markowitz}
\affiliation{Florida International University, University Park, Florida 33199}
\author{P.~McKee}
\affiliation{University of Virginia, Charlottesville, Virginia 22901}
\author{D.G.~Meekins}
\affiliation{Thomas Jefferson National Accelerator Facility, Newport News, Virginia 23606}
\author{H.~Mkrtchyan}
\affiliation{Yerevan Physics Institute, Yerevan, Armenia}
\author{B.~Moziak}
\affiliation{Rensselaer Polytechnic Institute, Troy, New York 12180}
\author{T.~Navasardyan}
\affiliation{Yerevan Physics Institute, Yerevan, Armenia}
\author{G.~Niculescu}
\affiliation{University of Virginia, Charlottesville, Virginia 22901}
\author{I.~Niculescu}
\affiliation{James Madison University, Harrisonburg, Virginia 22807}
\author{A.K.~Opper}
\affiliation{Ohio University, Athens, Ohio 45071}
\author{T.~Ostapenko}
\affiliation{Gettysburg College, Gettysburg, Pennsylvania 18103}
\author{P.E.~Reimer}
\affiliation{Argonne National Laboratory, Argonne, Illinois 60439}
\author{J.~Reinhold}
\affiliation{Florida International University, University Park, Florida 33199}
\author{J.~Roche}
\affiliation{Thomas Jefferson National Accelerator Facility, Newport News, Virginia 23606}
\author{S.E.~Rock}
\affiliation{University of Massachusetts Amherst, Amherst, Massachusetts 01003}
\author{E.~Schulte}
\affiliation{Argonne National Laboratory, Argonne, Illinois 60439}
\author{E.~Segbefia}
\affiliation{Hampton University, Hampton, Virginia 23668}
\author{C.~Smith}
\affiliation{University of Virginia, Charlottesville, Virginia 22901}
\author{G.R.~Smith}
\affiliation{Thomas Jefferson National Accelerator Facility, Newport News, Virginia 23606}
\author{P.~Stoler}
\affiliation{Rensselaer Polytechnic Institute, Troy, New York 12180}
\author{V.~Tadevosyan}
\affiliation{Yerevan Physics Institute, Yerevan, Armenia}
\author{L.~Tang}
\affiliation{Hampton University, Hampton, Virginia 23668}
\affiliation{Thomas Jefferson National Accelerator Facility, Newport News, Virginia 23606}
\author{V.~Tvaskis}
\affiliation{Nationaal Instituut voor Subatomaire Fysica, Amsterdam, The Netherlands}
\author{M.~Ungaro}
\affiliation{University of Connecticut, Storrs, Connecticut 06269}
\author{A.~Uzzle}
\affiliation{Hampton University, Hampton, Virginia 23668}
\author{S.~Vidakovic}
\affiliation{University of Regina, Regina, Saskatchewan, Canada, S4S 0A2}
\author{A.~Villano}
\affiliation{Rensselaer Polytechnic Institute, Troy, New York 12180}
\author{W.F.~Vulcan}
\affiliation{Thomas Jefferson National Accelerator Facility, Newport News, Virginia 23606}
\author{M.~Wang}
\affiliation{University of Massachusetts Amherst, Amherst, Massachusetts 01003}
\author{G.~Warren}
\affiliation{Thomas Jefferson National Accelerator Facility, Newport News, Virginia 23606}
\author{F.R.~Wesselmann}
\affiliation{University of Virginia, Charlottesville, Virginia 22901}
\author{B.~Wojtsekhowski}
\affiliation{Thomas Jefferson National Accelerator Facility, Newport News, Virginia 23606}
\author{S.A.~Wood}
\affiliation{Thomas Jefferson National Accelerator Facility, Newport News, Virginia 23606}
\author{C.~Xu}
\affiliation{University of Regina, Regina, Saskatchewan, Canada, S4S 0A2}
\author{L.~Yuan}
\affiliation{Hampton University, Hampton, Virginia 23668}
\author{X.~Zheng}
\affiliation{Argonne National Laboratory, Argonne, Illinois 60439}
\author{H.~Zhu}
\affiliation{University of Virginia, Charlottesville, Virginia 22901}

\date{\today}

\begin{abstract}
The differential cross-section for the process $p(e,e'p)\eta$ has been measured at $Q^2\protect\sim$  5.7 and 7.0 (GeV/c)$^2$ for centre-of-mass energies from threshold to 1.8 GeV, encompassing the $S_{11}(1535)$ resonance, which dominates the channel.  This is the highest momentum transfer measurement of this exclusive process to date.  The helicity-conserving transition amplitude $A_{1/2}$, for the production of the $S_{11}(1535)$ resonance, is extracted from the data.  Within the limited $Q^2$ now measured, this quantity appears to begin scaling as $Q^{-3}$---a predicted, but not definitive, signal of the dominance of perturbative QCD, at $Q^2\sim5$ (GeV/c)$^{2}$.
\end{abstract}



\maketitle

\ifthenelse{\equal{\twocol}{true}}{
	\newcommand{\includegraphicstwowidths}[3]{
		\includegraphics[width=#2\textwidth]{#1}
	}
}{
	\newcommand{\includegraphicstwowidths}[3]{
		\includegraphics[width=#3\textwidth]{#1}
	}
}

\section{Introduction}

The goal of strong interaction physics is to understand hadrons in terms of their fundamental constituents, the quarks and gluons.  Although these constituents are described by Quantum Chromodynamics (QCD), and perturbative methods work well where applicable, mostly the complexity of the theory precludes a description of hadrons in terms of QCD.  Various techniques are used to make progress, such as numerical simulation of QCD and hadron models with effective QCD degrees of freedom.  In this sense, the role of experiment is to make measurements which test the predictions of QCD-inspired quark models.  Most models can describe the static nucleon properties and the baryon spectrum, and so other measurements, such as electromagnetic transition form factors and strong decay amplitudes, are required.

A baryon's quark substructure can be excited into a resonance---an excited state of the quarks with well-defined baryon quantum numbers.  The transition form factor is the coupling (amplitude for the transition) from one baryon state to another, as a function of the squared invariant momentum transferred to the baryon $Q^2$.  The measurements of couplings between baryon states and the dependence of these on $Q^2$, can be used as stringent tests of quark models.  These couplings can be expressed in terms of the transition matrix elements between states of definite helicity.  

 The difficulty in measuring baryon transition form factors lies in isolating any of the multitude of wide and overlapping resonant states.  The $S_{11}(1535)$ is a baryon resonance that can be accessed relatively easily.   Although there are many overlapping states in its mass region, it is very strongly excited over the accessible $Q^2$ range and is the only resonance with a large branching fraction to $\eta$ mesons~\cite{Yao:2006px}, causing it to dominate the $p(e,e'p)\eta$ channel.  This dominance is partly due to isospin conservation, since the proton has isospin=$\frac{1}{2}$ and the $\eta$ has isospin=0, only the $N^{*} (I=\frac{1}{2})$ resonances can decay to a proton-$\eta$ final state---$N^{*}(I=\frac{3}{2})$ resonances are forbidden.  

As well as being accessible, the $S_{11}$ is an interesting resonant state.  It is the negative parity partner of the nucleon, they are both spin-half and isospin-half particles.  The transition form factor for the production of the $S_{11}$ falls more slowly with $Q^2$ than the dipole form factor $G_{D} = (1 + Q^2/0.71)^{-2}$, at least up to $Q^2 = 3.6$ GeV$^2$~\cite{armstrong99}, and more slowly than the form factor for typical baryons.  An example is the $D_{13}(1520)$~\cite{Brasse:1977as}, which is from the same $SU(6)\bigotimes O(3)$ multiplet and mass region as the $S_{11}(1535)$.  The $S_{11}(1535)$ branching fraction to $p\eta$, at $b_{\eta}\sim50\%$, is anomalously high when compared to that of the other $N^{*}$ resonances, a phenomenon which is not well understood.

It is expected from helicity conservation in perturbative QCD (pQCD) that at sufficiently high $Q^2$ the photocoupling amplitude $A_{1/2}$ will begin to scale as $1/Q^3$~\cite{Carlson:1988gt}, or equivalently the quantity $Q^3A_{1/2}$ will flatten.  The observation of such scaling is thus a possible signal of the transition to the dominance of hard processes.  This motivates the present experiment which studied exclusive $\eta$ production, allowing access to the amplitude $A_{1/2}$ for the $S_{11}$ resonance, at the highest ever $Q^2$ yet measured.

The first measurement of $\eta$ production at substantial $Q^2$ was published by Brasse \emph{et al.}~\cite{brasse84} in 1984 based on work at DESY that went to $Q^2=2.0$ and 3.0 GeV$^2$.  This was the first indication that the $S_{11}(1535)$ falls far slower with $Q^2$ than the $D_{13}(1520)$, and hence dominates the channel at high $Q^2$.

In 1999, Armstrong \emph{et al.}~\cite{armstrong99} published data obtained in Hall C at Jefferson Lab at $Q^2=2.4$ and 3.6 GeV$^2$, the highest until this work.   The cross-section was found to be about 30\% lower than the DESY data and the full width of the $S_{11}(1535)$ about twice as wide.  By comparing with inclusive data, a lower bound was put on the branching fraction $S_{11}\rightarrow\eta p$ of $b_{\eta}>0.45.$

A recent paper by the CLAS collaboration from Hall B at Jefferson Lab~\cite{Denizli:2007tq}, published data for this process, at centre-of-mass (c.m.) energy $W=1.5-2.3$ GeV and  $Q^2=0.13-3.3 \mathrm{~GeV}^2$.  The photocoupling amplitude $A_{1/2}$ of the proton to $S_{11}$(1535) transition was extracted, and the anisotropies in the differential cross-section were more precisely determined.  The results for the magnitude and width of the $S_{11}$ resonance favoured the Armstrong data over the older Brasse result.  Evidence was shown for a significant contribution to $\eta$ electroproduction due to a $P$-wave resonance with a mass around 1.7 GeV.

This paper describes an experiment where electrons were scattered off free protons at high momentum transfer and both electron and proton were detected in coincidence.  In Sec.~\ref{sec.formal} the kinematics of the reaction are discussed along with the formalities of the cross section and helicity amplitude for $S_{11}(1535)$ production.  Sec.~\ref{sec.exp} describes the apparatus and methods used to acquire the data.  Sec.~\ref{sec.data} then goes on to present the processing of the data including corrections, calibration, cuts, Monte Carlo simulation, backgrounds and ultimately the cross section extraction and error analysis.  In Sec.~\ref{sec.results} the $\eta$ production differential cross section is plotted and fit with an angular dependence.  A Breit-Wigner form is fitted to the data and the $S_{11}$ helicity amplitude and resonance parameters are extracted.  A brief summary is given in Sec.~\ref{sec.conc}.  The appendix tabulates the extracted $\eta$ production differential cross section.

\section{Formalism}
\label{sec.formal}

\subsection{Kinematics}

\begin{figure}[!htb]
\begin{center}
\includegraphicstwowidths{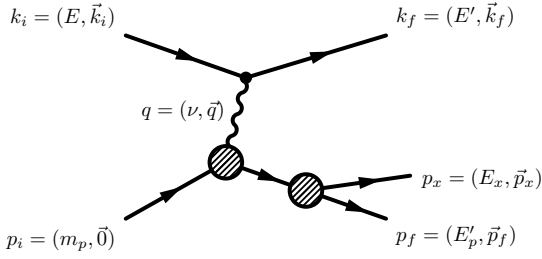}{0.48}{0.8}
\caption[one-photon exchange diagram for resonance electroproduction]{The one-photon exchange diagram of the resonance electroproduction process, where, for example, $k_i$ is the four-momentum vector of the incoming electron composed of energy $E$ and momentum $\vec{k}_i$.}\label{dis}
\end{center}
\end{figure}

Figure~\ref{dis} shows the one-photon exchange (Born) diagram for the resonance electroproduction process.  The incident electron $k_i$ scatters off the stationary proton $p_i$ with mass $m_p$.  We detect the scattered electron $k_f$ and proton $p_f$ and reconstruct the undetected particle $p_x$ using the missing mass technique, evaluated from four-momentum conservation
\begin{equation}
m_x^2=(p_i+k_i-p_f-k_f)^2.
\label{eqn.mmass2}
\end{equation}\\
Using the symbols from the diagram and neglecting the electron mass, the positive square of the four-momentum transferred from the lepton to hadron system is $Q^2\equiv -q^2=4EE'\mathrm{sin}^2(\theta_e/2)$.  The mass of the resonant state is $W^2=(q+p_i)^2=q^2+m_p^2+2m_p\nu$.  

\begin{figure}[!hbtp]
\begin{center}
\includegraphicstwowidths{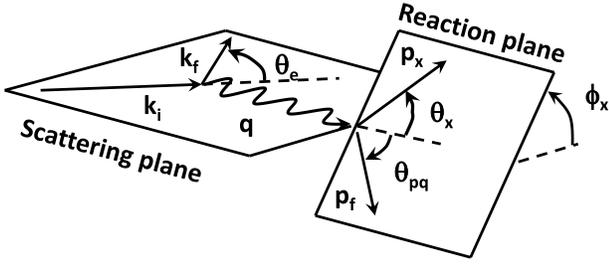}{0.48}{0.8}
\caption[Reaction plane coordinates~\cite{armstrong99}]{The scattering and reaction plane coordinate systems.}
\label{coordinates}
\end{center}
\end{figure}

Figure~\ref{coordinates} shows the scattering and reaction plane coordinate systems: $\theta_{e}$ is the scattering angle of the electron; $\theta_{pq}$ the angle between the outgoing proton and the momentum vector of the virtual photon, $\bm{q}$; the polar and azimuthal angles of the missing momentum are  $\theta^{*}_{x}$ and $\phi_{x}$ respectively, defined with respect to $\bm{q}$ and the electron scattering plane.  A super-scripted * denotes measurement in the $p\eta$ centre-of-momentum frame.

\subsection{Cross Section}

The five-fold differential cross-section for the reaction may be expressed as the product of the transverse virtual photon flux $\Gamma_{T}$ and the centre-of-mass cross-section for the electroproduction of the $p{\eta}$ pair
\begin{equation}
\frac{d^4\sigma}{dWdQ^2d\phi_ed\Omega^*_\eta} = \Gamma_{T}(W,Q^2)\frac{d\sigma}{d\Omega^*_\eta}(\gamma_{v} p\rightarrow p\eta), \\
\end{equation}
where the flux of transverse virtual photons in the Hand convention~\cite{hand63} is 
\begin{equation}
\Gamma_{T}(W,Q^2)=\frac{\alpha}{4\pi^2}\frac{W}{m_pE^2}\frac{K}{Q^2}\frac{1}{1-\epsilon}, \\
\end{equation}
the longitudinal polarisation of the virtual photon is given by
\begin{equation}
\epsilon=\frac{1}{1+2\frac{|\bm{q}|^2}{Q^2}\mathrm{tan}^2(\theta_e/2)}, \\
\end{equation}
and the energy required by a real photon to excite a proton to a resonance of mass $W$ is
\begin{equation}
\label{Keq}
K=\frac{W^2-m_p^2}{2m_p}.
\end{equation}

The unpolarised virtual photon cross-section is written in terms of the transverse polarised virtual photon $d\sigma_T/d\Omega^*_\eta$, longitudinal polarised virtual photon ${d\sigma_L}/{d\Omega^*_\eta}$ and interference contributions, ${d\sigma_{LT}}/{d\Omega^*_\eta}$ and ${d\sigma_{TT}}/{d\Omega^*_\eta}$ in Eq.~(\ref{diffcsexpand}).  Each of these four individual components are expressed in terms of multipoles~\cite{walker69,Warns:1989ie}, where $E_{l\pm}$, $M_{l\pm}$ and $S_{l\pm}$ are the electric, magnetic and scalar multipoles respectively; $l$ is the orbital angular momentum, and $\pm$ indicates the total angular momentum via $j=l\pm\frac{1}{2}$.  If only terms with $l\leq 2$ and either of the dominant isotropic multipoles, $E_{0+}$ or $S_{0+}$, are retained, then Eqs.~(\ref{diffcsmultipole}) are obtained~\cite{Knochlein:1995qz}.

The virtual photon cross-section is parametrised in terms of its angular dependence as Eq.~(\ref{eqn.multipole2}).  The parameters $A-F$ are then given in terms of the truncated multipole expansion by the Eqs.~(\ref{eqn.parapole}).

\ifthenelse{\equal{\twocol}{true}}{
\begin{widetext}
	\begin{eqnarray}
	\label{diffcsexpand}
	\frac{d\sigma}{d\Omega^*_\eta}(\gamma_{v} p\rightarrow p\eta) = \frac{d\sigma_T}{d\Omega^*_\eta} + \epsilon\frac{d\sigma_L}{d\Omega^*_\eta} + \sqrt{2\epsilon(1+\epsilon)}\frac{d\sigma_{LT}}{d\Omega^*_\eta}~\mathrm{cos}\phi^*_{\eta} +\epsilon\frac{d\sigma_{TT}}{d\Omega^*_\eta}~\mathrm{cos}2\phi^*_{\eta}
	\end{eqnarray}

\begin{eqnarray}
\frac{d\sigma_T}{d\Omega^*_\eta} & = & \frac{|\textbf{p}_{\eta}^*|W}{m_pK}\bigg\{|E_{0+}|^2 - \mathrm{Re}\Big[E_{0+}^*\big\{2\mathrm{cos}\theta_{\eta}^*M_{1-} -(3\mathrm{cos}^2\theta_{\eta}^*-1)(E_{2-}-3M_{2-})\big\}\Big]\bigg\};\nonumber\\
\frac{d\sigma_L}{d\Omega^*_\eta} & = & \frac{Q^2}{|\textbf{q}^*|^2}\frac{|\textbf{p}_{\eta}^*|W}{m_pK}\bigg\{|S_{0+}|^2 + 2\mathrm{Re}\Big[S_{0+}^*\big\{2\mathrm{cos}\theta_{\eta}^*S_{1-}  -2(1-3\mathrm{cos}^2\theta_{\eta}^*)S_{2-}\big\}\Big]\bigg\};\nonumber\\
\frac{d\sigma_{TL}}{d\Omega^*_\eta} & = & \sqrt{\frac{Q^2}{|\textbf{q}^*|^2}}\frac{|\textbf{p}_{\eta}^*|W}{m_pK}\bigg\{-\mathrm{sin}\theta_{\eta}^* \mathrm{Re}\Big[E_{0+}^*(S_{1-}+6\mathrm{cos}\theta_{\eta}^*S_{2-}) +S_{0+}^*\big\{M_{1-}+3\mathrm{cos}\theta_{\eta}^*(M_{2-}-E_{2-})\big\}\Big]\bigg\};  \nonumber\\
\frac{d\sigma_{TT}}{d\Omega^*_\eta} & = & \frac{|\textbf{p}_{\eta}^*|W}{m_pK}\bigg\{-3\mathrm{sin}\theta_{\eta}^* \mathrm{Re}\Big[E_{0+}^*(M_{2-}+E_{2-})\Big]\bigg\};
\label{diffcsmultipole}
\end{eqnarray}

\begin{equation}
\label{eqn.multipole2}
\frac{d\sigma}{d\Omega^*}  =  A + B~\mathrm{cos}\theta^* + C~\mathrm{cos}^2\theta^* + D~\mathrm{sin}\theta^*\mathrm{cos}\phi^* + E~\mathrm{cos}\theta^*\mathrm{sin}\theta^*\mathrm{cos}\phi^* + F~\mathrm{sin}^2\theta^*\mathrm{cos}2\phi^*
\end{equation}

\begin{eqnarray}
\label{eqn.parapole}
A & = & \frac{|\textbf{p}_{\eta}^*|W}{m_pK}\bigg\{|E_{0+}|^2 + \epsilon\frac{Q^2}{|\textbf{q}^*|^2}|S_{0+}|^2 -\Big(\mathrm{Re}\big[E_{0+}^*(E_{2-}-3M_{2-})\big] + 4\epsilon\frac{Q^2}{|\textbf{q}^*|^2}\mathrm{Re}\big[S_{0+}^*S_{2-}\big]\Big)\bigg\}  \nonumber\\
B & = & \frac{|\textbf{p}_{\eta}^*|W}{m_pK}\bigg\{-2\mathrm{Re}\big[E_{0+}^*M_{1-}\big]  + 2\epsilon\frac{Q^2}{|\textbf{q}^*|^2}\mathrm{Re}\big[S_{0+}^*S_{1-}\big]\bigg\}  \nonumber\\
C & = & \frac{|\textbf{p}_{\eta}^*|W}{m_pK}\bigg\{3\Big(\mathrm{Re}\big[E_{0+}^*(E_{2-}-3M_{2-})\big] + 4\epsilon\frac{Q^2}{|\textbf{q}^*|^2}\mathrm{Re}\big[S_{0+}^*S_{2-}\big]\Big)\bigg\}  \nonumber\\
D & = & \frac{|\textbf{p}_{\eta}^*|W}{m_pK}\bigg\{-\sqrt{2\epsilon(\epsilon+1)}\sqrt{\frac{Q^2}{|\textbf{q}^*|^2}}\mathrm{Re}\big[E_{0+}^*S_{1-}+S_{0+}^*M_{1-}\big]\bigg\}  \nonumber\\
E & = & \frac{|\textbf{p}_{\eta}^*|W}{m_pK}\bigg\{-3\sqrt{2\epsilon(\epsilon+1)}\sqrt{\frac{Q^2}{|\textbf{q}^*|^2}}\mathrm{Re}\big[2E_{0+}^*S_{2-}+S_{0+}^*(M_{2-}-E_{2-})\big]\bigg\}  \nonumber\\
F & = & \frac{|\textbf{p}_{\eta}^*|W}{m_pK}\Big\{-3\epsilon \mathrm{Re}\big[E_{0+}^*(E_{2-}+M_{2-})\big]\Big\} 
\end{eqnarray}
\end{widetext}

}{
	\begin{eqnarray}
	\label{diffcsexpand}
	\frac{d\sigma}{d\Omega^*_\eta}(\gamma_{v} p\rightarrow p\eta) = & \sigma_T + \epsilon\sigma_L + \sqrt{2\epsilon(1+\epsilon)}\sigma_{LT}~\mathrm{cos}\phi^*_{\eta} \nonumber\\
	& +\epsilon\sigma_{TT}~\mathrm{cos}2\phi^*_{\eta}
	\end{eqnarray}

\begin{eqnarray}
\frac{d\sigma_T}{d\Omega^*_\eta} & = & \frac{|\textbf{p}_{\eta}^*|W}{m_pK}\bigg\{|E_{0+}|^2 - \mathrm{Re}\Big[E_{0+}^*\big\{2\mathrm{cos}\theta_{\eta}^*M_{1-} \nonumber\\ 
	& & -(3\mathrm{cos}^2\theta_{\eta}^*-1)(E_{2-}-3M_{2-})\big\}\Big]\bigg\};\nonumber\\
\frac{d\sigma_L}{d\Omega^*_\eta} & = & \frac{Q^2}{|\textbf{q}^*|^2}\frac{|\textbf{p}_{\eta}^*|W}{m_pK}\bigg\{|S_{0+}|^2 + 2\mathrm{Re}\Big[S_{0+}^*\big\{2\mathrm{cos}\theta_{\eta}^*S_{1-} \nonumber\\ 
	& & -2(1-3\mathrm{cos}^2\theta_{\eta}^*)S_{2-}\big\}\Big]\bigg\};\nonumber\\
\frac{d\sigma_{TL}}{d\Omega^*_\eta} & = & \sqrt{\frac{Q^2}{|\textbf{q}^*|^2}}\frac{|\textbf{p}_{\eta}^*|W}{m_pK}\bigg\{-\mathrm{sin}\theta_{\eta}^* \mathrm{Re}\Big[E_{0+}^*(S_{1-}+6\mathrm{cos}\theta_{\eta}^*S_{2-})   \nonumber\\ 
	& & +S_{0+}^*\big\{M_{1-}+3\mathrm{cos}\theta_{\eta}^*(M_{2-}-E_{2-})\big\}\Big]\bigg\};  \nonumber\\
\frac{d\sigma_{TT}}{d\Omega^*_\eta} & = & \frac{|\textbf{p}_{\eta}^*|W}{m_pK}\bigg\{-3\mathrm{sin}\theta_{\eta}^* \mathrm{Re}\Big[E_{0+}^*(M_{2-}+E_{2-})\Big]\bigg\};
 \label{diffcsmultipole}
\end{eqnarray}

\begin{eqnarray}
\label{eqn.multipole2}
\frac{d\sigma}{d\Omega^*} & = & A + B~\mathrm{cos}\theta^* + C~\mathrm{cos}^2\theta^* + D~\mathrm{sin}\theta^*\mathrm{cos}\phi^* \nonumber\\ 
&& + E~\mathrm{cos}\theta^*\mathrm{sin}\theta^*\mathrm{cos}\phi^* + F~\mathrm{sin}^2\theta^*\mathrm{cos}2\phi^*
\end{eqnarray}

\begin{eqnarray}
\label{eqn.parapole}
A & = & \frac{|\textbf{p}_{\eta}^*|W}{m_pK}\bigg\{|E_{0+}|^2 + \epsilon\frac{Q^2}{|\textbf{q}^*|^2}|S_{0+}|^2  \nonumber\\
	& & -\Big(\mathrm{Re}\big[E_{0+}^*(E_{2-}-3M_{2-})\big] + 4\epsilon\frac{Q^2}{|\textbf{q}^*|^2}\mathrm{Re}\big[S_{0+}^*S_{2-}\big]\Big)\bigg\}  \nonumber\\
B & = & \frac{|\textbf{p}_{\eta}^*|W}{m_pK}\bigg\{-2\mathrm{Re}\big[E_{0+}^*M_{1-}\big]  + 2\epsilon\frac{Q^2}{|\textbf{q}^*|^2}\mathrm{Re}\big[S_{0+}^*S_{1-}\big]\bigg\}  \nonumber\\
C & = & \frac{|\textbf{p}_{\eta}^*|W}{m_pK}\bigg\{3\Big(\mathrm{Re}\big[E_{0+}^*(E_{2-}-3M_{2-})\big] \nonumber\\
	& & + 4\epsilon\frac{Q^2}{|\textbf{q}^*|^2}\mathrm{Re}\big[S_{0+}^*S_{2-}\big]\Big)\bigg\}  \nonumber\\
D & = & \frac{|\textbf{p}_{\eta}^*|W}{m_pK}\bigg\{-\sqrt{2\epsilon(\epsilon+1)}\sqrt{\frac{Q^2}{|\textbf{q}^*|^2}}\mathrm{Re}\big[E_{0+}^*S_{1-}\nonumber\\
	& & +S_{0+}^*M_{1-}\big]\bigg\}  \nonumber\\
E & = & \frac{|\textbf{p}_{\eta}^*|W}{m_pK}\bigg\{-3\sqrt{2\epsilon(\epsilon+1)}\sqrt{\frac{Q^2}{|\textbf{q}^*|^2}}\mathrm{Re}\big[2E_{0+}^*S_{2-}\nonumber\\
	& & +S_{0+}^*(M_{2-}-E_{2-})\big]\bigg\}  \nonumber\\
F & = & \frac{|\textbf{p}_{\eta}^*|W}{m_pK}\Big\{-3\epsilon \mathrm{Re}\big[E_{0+}^*(E_{2-}+M_{2-})\big]\Big\} 
\end{eqnarray}

}

\subsection{Helicity Amplitude}
\label{sec.helicity}

The helicity amplitude is the matrix element that connects states of definite (the same or different) helicity.  As such, it is a convenient measure of the coupling strength between states and can be used to fundamentally test quark models.  The amplitudes are labelled by the virtual photon polarisation (either transverse $A$ or longitudinal $S$) and the total $\gamma N$ helicity ($\frac{1}{2}$ or $\frac{3}{2}$).  Spin-$\frac{1}{2}$ resonances are therefore described only by $A_{1/2}$ and $S_{1/2}$.  

The helicity amplitude $A_{1/2}$, for the process $\gamma_vp \rightarrow S_{11}(1535)$, can be obtained from the contribution of the $S_{11}(1535)$ to the $E_{0+}$ multipole at the resonant mass $W=W_R$, using~\cite{Drechsel:1992pn,PDG76}
\begin{equation}
\label{eqn.helicity2multipole}
A_{1/2}=\left[2\pi\frac{|\bm{p}_{\eta}^*|_RW_R}{m_pK}\frac{W_R}{m_p}\frac{\Gamma_R}{b_{\eta}}\right]^{1/2}|E_{0+}(W_R)|.
\end{equation}

This requires, not only isolating the $S_{11}(1535)$ from the other resonances and the non-resonant background, but further isolating the $E_{0+}$ multipole from the other multipoles.  In this case, for $\gamma_vp \rightarrow \eta p$ at the $S_{11}(1535)$ resonance mass, such an isolation is almost implicit in the measurement due to the dominance of the $S_{11}(1535)$.  Being an $S$-wave resonance, implies a dominance of the isotropic multipoles---which has previously been seen in the data~\cite{armstrong99,krusche95,brasse84,Denizli:2007tq}.  So too, among the isotropic contributions it appears that the transverse multipole $E_{0+}$, dwarfs the longitudinal part $S_{0+}$~\cite{Brasse:1977as,Breuker:1978qr,Denizli:2007tq}.

Doing a longitudinal/transverse ($LT$) separation requires measuring the cross-section for at least two values of $\epsilon$ at the same $Q^2$, which was not done in this experiment.  Such separations performed in the late 1970's are consistent with no longitudinal component.  Where $R = \sigma_L/\sigma_T$ is the ratio of longitudinal to transverse cross sections,  Ref~\cite{Breuker:1978qr} found $R$ = 0.23 $\pm$ 0.15 at $Q^2$ = 0.4 GeV$^2$ and Ref~\cite{Brasse:1977as} found $R$ = 0.22 $\pm$ 0.23 at $Q^2$ = 0.6 GeV$^2$ and $R$ = -0.16 $\pm$ 0.16 at $Q^2$ = 1 GeV$^2$.  Quark models~\cite{Ravndal:1972pn} show this ratio decreasing with $Q^2$.  In angular fit to their recent data, Denizli~\cite{Denizli:2007tq} shows that the parameters of the $P^1(\mathrm{cos}\theta^*_{\eta})\mathrm{cos}\phi$ components fluctuate around zero and are consistent with zero within experimental uncertainty.  These parameters measure the $d\sigma_{LT}/d\Omega$ component of $d\sigma/d\Omega$, suggesting that the longitudinal component is small---but since ${d\sigma_{\mathrm{LT}}}/{d\Omega^*}$  is a sum of terms with possibly different signs, it is possible that $S_{0+}$ is in fact comparable to $E_{0+}$.  

In this paper it is assumed that the longitudinal amplitudes are not significant for this reaction.  The validity of this will become clear in the future when $LT$ separations are done at high $Q^2$.

The cross-section can thus be written as depending only on the dominant $E_{0+}$ multipole in the simple form 
\begin{equation}
\label{eqn.cs2multipole}
\frac{d\sigma}{d\Omega^*_{\eta}} \approx \frac{|\bm{p}_{\eta}^*|W}{m_pK}|E_{0+}|^2.
\end{equation}
The combination of Eq.~(\ref{eqn.cs2multipole}) and Eq.~(\ref{eqn.helicity2multipole}) yields 
\begin{equation}
A_{1/2}(Q^2)=\sqrt{\frac{W_R\Gamma_R}{2m_pb_{\eta}}\sigma_R(Q^2)},
\label{eqn.helicity2cs}
\end{equation}
the helicity amplitude as a function of $\sigma_{R} \equiv \sigma(W_R)$ [the total cross-section of the $S_{11}(1535)$ resonance, measured at the resonance mass $W_R$.]

The $E_{0+}$ multipole can be more reliably extracted from a fit to the angular dependence.    Parameters $A$ and $C$ in Eqs.~(\ref{eqn.parapole}) share some common terms, and a simple cancellation yields Eq.~(\ref{eqn.ApCo3})---although in the absence of an $LT$ separation, it still must be assumed that $S_{0+}$ is negligible.
\begin{eqnarray}
A + \frac{1}{3}C & = & \frac{|\textbf{p}_{\eta}^*|W}{m_pK}\bigg\{|E_{0+}|^2 + \epsilon\frac{Q^2}{|\textbf{q}^*|^2}|S_{0+}|^2 \bigg\} \nonumber\\
 		& \approx & \frac{|\textbf{p}_{\eta}^*|W}{m_pK}|E_{0+}|^2
\label{eqn.ApCo3}
\end{eqnarray}

Where possible in this work, the $E_{0+}$ multipole is extracted using both methods, but for consistency with previous analyses the final result is quoted from the method assuming isotropy.

\section{The Experiment}
\label{sec.exp}

\begin{figure}[hbt]
\begin{center}
\includegraphicstwowidths{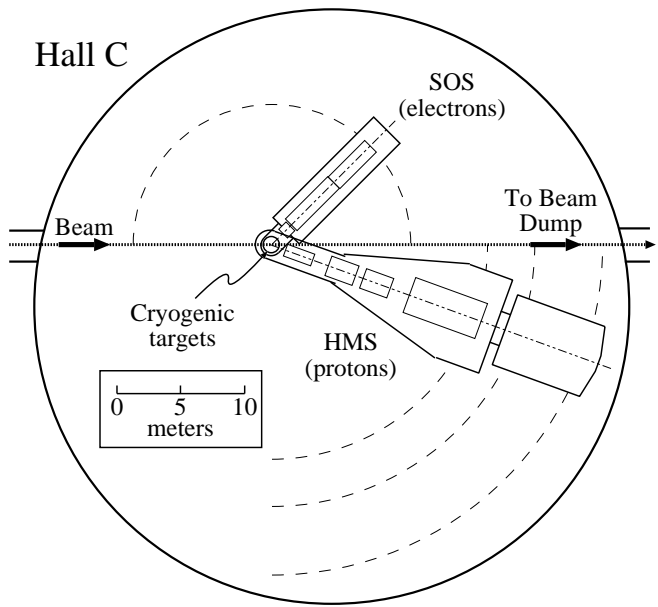}{0.48}{0.8}
\caption[Plan view of Hall C~\cite{armstrong99}]{A plan view of Hall C showing the beamline, target and the SOS and HMS spectrometers which detected electrons and protons respectively.  Figure from Ref.~\cite{armstrong99}.}
\label{fig.HallC}
\end{center}
\end{figure}

The experiment, measuring the unpolarised differential cross-section for the process $p(e,e'p)\eta$, was performed in Hall C (Fig.~\ref{fig.HallC}) of the Thomas Jefferson National Accelerator Facility, during May and June of 2003.  The Short Orbit Spectrometer (SOS)~\cite{Mohringphd}, a resistive $QD\overline{D}$ (quadrupole, dispersive dipole, anti-dispersive dipole) spectrometer, was used to detect scattered electrons.  The High Momentum Spectrometer (HMS)~\cite{Arringtonphd}, with a superconducting $QQQD$ configuration, detected the recoil protons.  The $\eta$ particles were identified using the missing mass method.

\begin{figure}[hbt]
\begin{center}
\includegraphicstwowidths{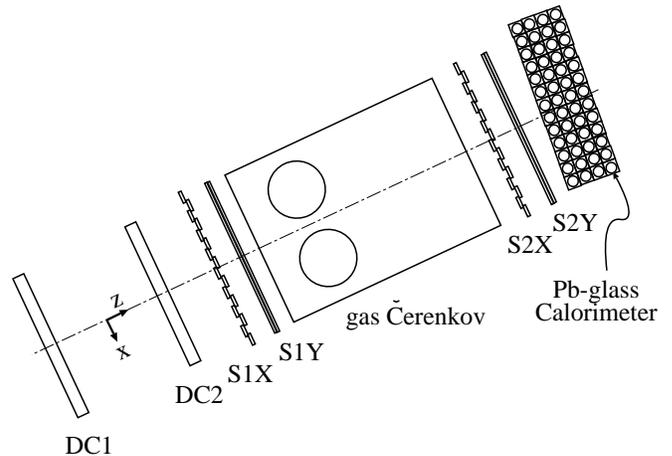}{0.48}{0.8}
\caption[Spectrometer detector elements, side view~\cite{armstrong99}]{A side view of the HMS detector stack, which is also representative of the SOS.  The detected particles travel from left to right, encountering first the two drift chambers (DC) then the first two arrays of scintillators (S1) oriented in the X and Y directions, then the gas \v Cerenkov detector, the third and fourth scintillator arrays and finally the calorimeter.  Figure from Ref.~\cite{armstrong99}}
\label{fig.detectors}
\end{center}
\end{figure}

Both spectrometers have a similar detector ensemble, including drift chambers for determining the track, scintillator arrays for triggering, an electromagnetic calorimeter for particle identification (PID) and a threshold gas \v{C}erenkov also for PID and tuned to differentiate between pions and electrons in the SOS.  Figure~\ref{fig.detectors}, showing the detector components, is representative of either detector stack.

The Jefferson Laboratory's superconducting radiofrequency Continuous Electron Beam Accelerator Facility (CEBAF) provides multi-GeV continuous-wave beams for experiments at the nuclear and particle physics interface~\cite{Leemann:2001dg}.  The accelerator consists of two anti-parallel linacs linked by nine recirculation beam lines in the shape of a racetrack, for up to five passes.  Beam energies up to nearly 6 GeV at 100 $\mu$A and $>75\%$ polarization are possible.  For this experiment, the incident electrons had the maximum available energy: $E_{e}$ = 5.500 GeV for most of the experiment and $E_{e}$ = 5.491 GeV for an 11 day period near the beginning.  

The target was liquid hydrogen maintained at a temperature of 19~K.  The beam passes through 3.941~cm of liquid and through 0.12~mm of aluminium target cell walls on entrance and exit.  The beam was rastered within a square of $\pm$ 1~mm to minimise density changes due to target boiling.  A dummy target consisting of two aluminium plates was used to simulate reactions within the target walls.

The trigger for the experiment was a coincidence between pre-triggers (or singles triggers) from both of the spectrometers.  Both of the spectrometer pre-triggers were the requirement of a signal in three out of the four scintillator planes (SCIN).  In addition to the coincidence trigger, data were taken for singles triggers from both of the two spectrometers.  This was pre-scaled according to the rate so as not to interfere with the coincidence trigger.  This singles data allowed the monitoring of the luminosity and the electron detection efficiency.  The elastic scattering events within the SOS were used to monitor the beam energy and the performance of the SOS magnets.

Blok \textit{et al.}~\cite{Blok:2008jy} is descriptive of the accelerator, beam monitoring equipment and current monitors, target rastering system, beam energy measurement and cryogenic target.  More detailed discussions are made of the two spectrometers, their detector packages, the trigger logic and data acquisition.  Further references are provided for all covered topics, the interested reader is advised to consult that work.

The electron spectrometer was fixed in angle and momentum, thereby defining a central three-momentum transfer vector $\vec{q}$ for the virtual photon which mediates the reaction.  Around this $\vec{q}$ vector is a cone of reaction products including the protons from the resonance decay of interest in this measurement.  The ``kinematic focusing'' caused by the high momentum transfer of the reaction makes it possible to capture a large fraction of centre of mass decay solid angle in a spectrometer, as it comes out as a ``narrow'' cone in the lab.  The proton spectrometer was stepped in overlapping angle and momentum steps to capture as much of this decay cone as possible. 

\begin{figure}[!hbtp]
\begin{center}
\includegraphicstwowidths{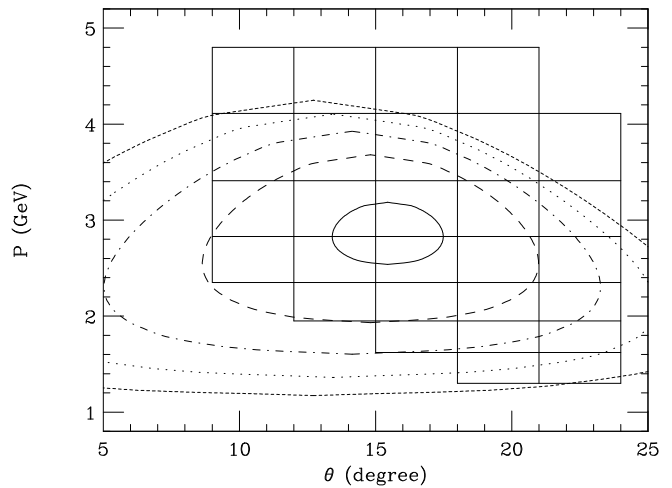}{0.48}{0.9}
\caption[Angle and momentum acceptance, lower-$Q^2$]{The $W$ acceptance of the detector pair for the $e(p,e'p)\eta$ reaction, in the lower-$Q^2$ configuration, as a function of the laboratory scattering angle and momentum of the proton.  The contours are constant $W$ of the hadronic system, for an electron at 47.5 degrees and momentum of 1.74 GeV/c, for the full range of $\theta^*$ and $\phi=0$ and 180 degrees.  The solid central contour is $W$ = 1.5 GeV, from which they increase in steps of 100 MeV to the outermost at $W$ = 1.9 GeV.  In practice, the angle and momentum bite of the SOS causes the contours to be much broader.  Each black box is the acceptance of a particular HMS setting ($^\dag$Table~\ref{tab.kinmat}).  The alternate settings are offset by 1.5$^{\circ}$ and are a 4.7\% increase in momentum, so that they are approximately centred on the points where the boxes join.}
\label{fig.anglecover}
\end{center}
\end{figure}

The exact choice of kinematics was based on a compromise between maximising the $Q^2$ for the available beam energy and detecting the full centre-of-mass decay cone for the $p(e,e'p)\pi^0$ reaction to the highest possible $W$.  This reaction, which was measured concurrently is reported on by Villano~\cite{Villanophd}.  The maximum central momentum of the SOS, 1.74 GeV, required increasing $\theta_{\mathrm{SOS}}$ to increase the $Q^2$, while the minimum HMS angle of 10.5 degrees required decreasing $\theta_{\mathrm{SOS}}$ to extend the full angular coverage to higher $W$.  At $\theta_{\mathrm{SOS}}$ = 47.5 degrees and the maximum SOS momentum, it was found that the kinematic region from  pion threshold to above the $S_{11}$ mass fell nicely within the best resolution region of the SOS spectrometer and full cos$\theta^*$ coverage was possible for the $p\pi^0$ up to $W =$ 1.4 GeV and  $p\eta$ up to $W =$ 1.6 GeV, as shown in Fig.~\ref{fig.anglecover}.  These SOS central parameters correspond to a virtual photon with momentum 4.51~GeV and angle 16.5~degrees and $Q^2 \sim 5.8$ (GeV/$c$)$^2$ at the $S_{11}$ resonance mass.

In addition to data taken with these kinematics, it was decided to take a smaller set of data at even higher $Q^2$, although the angular coverage would be incomplete.  In this configuration the SOS was set with central momentum of 1.04~GeV and angle of 70~degrees, which gives a central virtual photon with $|\vec{q}|$ = 5.24~GeV and angle 10.8~degrees and $Q^2 \sim 7.0$ (GeV/$c$)$^2$ at the $S_{11}$ resonance mass.  For the purposes of this paper, the first data set will be called the `lower-$Q^2$' configuration and the second data set, the `higher-$Q^2$'.  The kinematic settings for the experiment are summarised in Table~\ref{tab.kinmat}.

\begin{table}[!hbt]
\begin{center}
\begin{tabular}{cc|cr}
\hline
\hline
\multicolumn{2}{c|}{Electron Arm} & \multicolumn{2}{c}{Proton Arm} \\
$p_{\mathrm{SOS}}$ & $\theta_{\mathrm{SOS}}$ & $p_{\mathrm{HMS}}$ 	& \multicolumn{1}{c}{$\theta_{\mathrm{HMS}}$}\\
GeV		& degrees		& GeV			& \multicolumn{1}{c}{degrees}		\\
\hline
	&		& 4.70\phantom{$^\dag$}		& 18.0, 15.0 \\
	&		& 4.50$^\dag$	& 19.5, 16.5, 13.5, 11.2 \\
	&		& 3.90\phantom{$^\dag$}  	& 21.0, 18.0, 15.0, 12.0 \\
	&		& 3.73$^\dag$	& 22.5, 19.5, 16.5, 13.5, 11.2\\
1.74 	& 47.5 		& 3.24\phantom{$^\dag$} 		& 24.0, 21.0, 18.0, 15.0, 12.0\\
	& 		& 3.10$^\dag$	& 22.5, 19.5, 16.5, 13.5, 11.2\\
	&		& 2.69\phantom{$^\dag$}		& 24.0, 21.0, 18.0, 15.0, 12.0\\
	&		& 2.57$^\dag$  	& 22.5, 19.5, 16.5, 13.5, 11.2\\
	&		& 2.23\phantom{$^\dag$}		& 21.0, 18.0, 15.0, 12.0 \\
	&		& 2.13$^\dag$	& 22.5, 19.5, 16.5, 13.5 \\
\hline
	&		& 4.70 & 11.2 \\
 	&		& 4.50 & 14.2 \\
1.04 	& 70.0 		& 3.90 & 11.2 \\
 	&		& 3.73 & 14.2, 11.2 \\
 	&		& 3.24 & 11.2 \\
\hline
\hline
\end{tabular}
\end{center}
\caption[Kinematics]{\label{tab.kinmat}The kinematic settings of the two spectrometers.}
\end{table}

Data were taken at a mean beam current of 92 $\mu$A.  The lower-$Q^2$ configuration was run for 6 weeks, totaling 127 C of electrons through the target from which about 50,000 $\eta$ particles were identified from proton-electron coincidences, by missing mass reconstruction.  Due to improved accelerator operation, the one week of running for the higher-$Q^2$ setting received 29 C of charge, but only about 2,000 $\eta$ particles were reconstructed.

\section{Data Analysis}
\label{sec.data}

The raw data as recorded by the electronics were replayed offline to produce PAW or \textsc{root} ntuples of calibrated physics quantities.  Corrections were made to the data for inefficiencies, dead times and accidental coincidences.   The detector response was simulated using the Monte Carlo technique (including multiple scattering in the detector and nuclear reactions in the target walls) with one input model cross-section for the $\eta$ production signal and another model for the multipion background processes, described in detail in Sec.~\ref{sec.montecarlo}.  Using an iterative procedure, a linear combination of the signal and background simulations was fitted to the data and the result used to refine the simulation input model, until the simulation in each bin matched the data with a multiplicative factor of close to unity.

\subsection{Raw Data to Physical Quantities}

The raw data from each trigger was stored onto tape.  These data were ``replayed'' offline a number of times during the analysis, using the Hall C data reduction code, as the calibration of the detectors was improved.  For each event, a list of calibrated event properties including position and angles of the track, timing and energy deposition information were determined.  So too were quantities for the scattering including the centre-of-mass angles, invariant hadronic mass and the missing mass.  For each run an ntuple of these event parameters was produced along with a file containing scaler information and calculated efficiencies and dead times for that run.  

The data were corrected on a run-by-run basis for these inefficiencies and dead times during the filling procedure---each event passing the cuts was filled into the histogram weighted by a run dependent correction factor.  This included track reconstruction inefficiencies in the HMS and SOS spectrometers and computer and electronic dead times.  A summary of all the corrections applied to the data is given in Table~\ref{tab.corr}.

\begin{table}[!hbtp]
\begin{center}
\begin{tabular}{lrlrl}
\hline
\hline
Effect				& \multicolumn{2}{c}{lower-$Q^2$}	& \multicolumn{2}{c}{higher-$Q^2$}  \\
\hline
Proton absorption 		& \multicolumn{4}{c}{$+4 \pm 1$\%} \\
$^{\dag}$Computer DT		& +(1.0 &$-$ 19.1)\%   & +(1.8 &$-$ 10.9)\%  \\
$^{\dag}$HMS tracking		& +(2.3 &$-$ 14.3)\%   & +(3.3 &$-$ ~7.4)\%   \\
$^{\dag}$SOS tracking 		& +(0.3 &$-$ ~0.9)\%   & +(0.2 &$-$ ~0.8)\%   \\
$^{\dag}$Electronics DT		& +(0.0 &$-$ ~2.4)\%   & +(0.0 &$-$ ~0.6)\%   \\
$^{\ddag}$Random coincidence	& $-$(0.0 &$-$ ~7.6)\% & $-$(0.0 &$-$ ~1.2)\%\\
\hline
\hline
\end{tabular}
\end{center}
\caption[Corrections to data]{\label{tab.corr}Corrections applied to the data.  For corrections applied $^{\dag}$run-by-run or $^{\ddag}$bin-by-bin, the range of the size is indicated in parentheses.}
\end{table}

The pion form factor ($F_{\pi}$) experiment~\cite{Hornphd,Blok:2008jy} was conducted in the same suite of experiments as the current experiment and this work makes reference to some analyses reported there.  A detailed description of the fitting of the reconstruction matrix elements for the spectrometers is included there.  A number of offsets and corrections were determined by analysing singles elastic scattering and coincident $^1H(e,e'p)$ events.  From these kinematically overdetermined reactions, it was possible to check the momentum $p$ and angles $\theta$ and $\phi$ (in-plane and out-of-plane relative to the spectrometer central axis respectively) for both spectrometers, and the beam energy $E$.  A fit was done to determine what offsets to these quantities most accurately produced the required values for the invariant hadronic mass, and missing mass and energy for the elastic scattering.  In the case of SOS momentum (equivalently the dipole field), there is a saturation as the current is increased due to the resistive nature of the magnets.  A field dependent correction was thus determined.  These offsets, summarised in Table~\ref{tab.fpicorr}, were used in the replay of the present data.

\begin{table}[!hbtp]
\begin{center}
\begin{tabular}{lr@{}c@{}lr@{}c@{}l}
\hline
\hline
Quantity	& \multicolumn{3}{c}{HMS}	& \multicolumn{3}{c}{SOS}  \\
\hline
$\theta$		& 0.0~ 		& $\pm$ & ~0.5 mrad	& 0.0~ 		& $\pm$ & ~0.5 mrad \\
$\phi$			& $+ 1.1$~	& $\pm$ & ~0.5 mrad	& $+3.2$~	& $\pm$	& ~0.5 mrad \\
$p$ (lower-$Q^2$)	& $- 0.13$~ 	& $\pm$ & ~0.05\% 	& $-1.36$~ 	& $\pm$	& ~0.05\% \\
$p$ (higher-$Q^2$)	& 		&	&		& $0.00$~	& $\pm$	& ~0.05\% \\
\hline
$E_{e}$		& \multicolumn{6}{c}{$0.00 \pm 0.05$\%} \\
\hline
\hline
\end{tabular}
\end{center}
\caption[Spectrometer offsets applied to the data]{\label{tab.fpicorr}Nominal 2003 spectrometer offsets~\cite{Hornphd,Blok:2008jy} applied to the data during the replay phase.}
\end{table}

\subsubsection{Trigger Efficiency}
\label{sec.trigeff}

The HMS trigger was a three out of four coincidence between the four scintillator planes.  A trigger inefficiency for proton detection in the HMS is produced by protons which are not detected in their interaction with the scintillator, and by protons that do not make it through all the scintillators due to absorption.

A previous study of general HMS trigger efficiency~\cite{Hornphd,Blok:2008jy} showed a strong dependence on relative particle momentum $\delta_{\mathrm{HMS}}$.  The momentum in the spectrometers is measured relative to the central momentum $p_{\mathrm{set}}$, so that particles with the same $\delta = (p-p_\mathrm{set})/p_{\mathrm{set}}$ are dispersed by the same amount.  The trigger efficiency was mostly very high at 0.995 but dropped rapidly for momenta lower than $\delta\sim-6\%$.  The data was analysed with a cut of $\delta>-6$ resulting in an average increase in extracted cross-section of 1.4\%.  No correction for this effect was made, but this figure was used as an estimate of the error due to the trigger efficiency.

The trigger requires hits in scintillator planes S1 and S2, so another source of inefficiency is absorption, through nuclear reactions, of the proton in target or detector materials before the S2 plane.  The total $pp$ collision cross-section, $\sigma_{pp}$, varies slightly from 47 to 42~mb for proton lab momenta between 2 to 5~GeV/$c$~\cite{Yao:2006px} which is the momentum range of this experiment.  Therefore for this experiment, the trigger efficiency due to absorption is relatively independent of kinematic setting.  

The primary sources of interacting material are the S1 scintillator planes which had a thickness of 1~cm each and the Aluminium windows of the gas \v Cerenkov and aerogel detectors which had a total thickness of 0.51~cm.  The proton-nuclear cross-section was estimated as  $A^{0.7}\sigma_{pp}$.  Combining interactions in all material, the trigger efficiency due to proton absorption is estimated to be 0.95.

To calculate the correction used in the experiment for the trigger efficiency due to proton absorption, a study of $ep$ elastic events was done.  The SOS was set for electrons at central angle = 50$^{\circ}$ and central momentum of 1.74~GeV/$c$ and the HMS was set for protons at central angle of 18$^{\circ}$ and central momentum of 4.34~GeV/$c$ at a beam energy of 5.247~GeV. For a point target, the SOS has an out-of-plane angular acceptance of $\pm$37~mr and an in-plane angular acceptance of $\pm$57~mr (the in and out-of-plane angles are relative to the central axis of the spectrometer), while the HMS has an out-of-plane angular acceptance of $\pm$70~mr and an in-plane angular acceptance of $\pm$27~mr.  The ratio of electron to proton momentum is 0.4, so for the maximum SOS out-of-plane angle, the corresponding HMS out-of-plane angle is 15~mr.  The maximum  SOS in-plane angle gives a corresponding HMS in-plane angle of 23~mr.  The data acquisition is set-up to accept singles triggers from the SOS and HMS individually in addition to  coincidence triggers between the HMS and SOS.

In offline analysis, the cuts described in Sec.~\ref{sec.sospid} were used to identify electrons in the SOS.  A good elastic event in the SOS was identified by a cut of 0.9 $< W <$ 1.0~GeV and a cut on the SOS in-plane angle of $\pm$50~mr which ensured that the proton would be within the HMS angular acceptance.   The proton was selected by the time-of-flight between electron and the particle detected in the HMS.  The raw number of coincident $ep$ events was 2009 and the number of single events was 205.  

Some data were taken with an aluminium  ``dummy'' target, which is intended to model an empty target cell, but is 7.78 times thicker in order to increase the count rate.  Analysis of this data determined that the target endcaps would contribute 12 $\pm$ 3 events to the raw coincidence events and 123.0 $\pm$ 12 events to the raw single events.  Therefore the proton trigger efficiency due to absorption is 0.96 $\pm$ 0.01, in good agreement with the prediction.  A 4\% correction was applied to the data for this effect. 

\subsubsection{Calibration of Simulation Resolution}
\label{sec.res}

As described in Section~\ref{sec.etaID}, the cross section is obtained by integrating over the $\eta$ missing mass peak.  The sensitivity to the exact shape of the peak is thus small and is minimised by matching the simulation resolution to the data resolution as far as possible.  This was done for the \emph{elastic} peak, following which the simulation of the eta peak was in agreement with the data width without further change.

The elastic scattering of electrons into the SOS and protons into the HMS were compared to SIMC Monte Carlo simulations of the same.  The invariant mass determined from elastic scattering must be the proton mass, but is broadened due to resolution effects and radiative tails---which are included in the simulation.  In both detectors it was found that the width of this peak predicted by the simulation was narrower than for the data.  These resolution differences were taken into account by increasing the drift chamber resolutions in each spectrometer, from the nominal value of 300~$\mu$m.  The resolution was varied, and the simulation repeated, until a Gaussian fitted to the simulated spectrum had the same width as a Gaussian (and polynomial background) fitted to the data.  It was found that the HMS needed a drift chamber resolution of 570~$\mu$m to match the data width of 17.2~MeV, while in the SOS, the 30.1~MeV width was achieved with a 350~$\mu$m resolution.

This method had the effect of degrading the optics of the simulation
slightly to match the experimental transport matrix elements of the
data, in a logical yet simple manner.  During the systematic error
analysis process, described in Section~\ref{sec.errors}, the drift
chamber resolutions were varied by 10\% and the integration limits
were varied by 0.1 GeV$^2$, characterising the sensitivity of the
extracted  differential cross-sections and amplitudes.

\subsubsection{Collimator Punch Through}

A source of background is due to particles that interact with the edges of the HMS collimator aperture, located just before the first quadrupole magnet, whose kinematics are thus changed.  The collimator is made from 6.35 cm thick HEAVYMET (machinable Tungsten with 10\% CuNi; density=17~$g/cm^{3}$.)  For practical purposes electrons are stopped by the SOS collimator, but protons have the possibility of ``punching'' through the collimator, undergoing multiple scattering and energy loss in the material, and still making it through the spectrometer to the detectors.  This process is modelled in the simulation of the experiment and additionally a loose cut, 3 centimeters outside the collimator edge, is used to eliminate unphysical reconstructions.  

\subsubsection{Data Cuts}

The `standard' cuts are listed in Table~\ref{tab:cuts}.  The cuts on relative electron momentum $\delta_{\mathrm{SOS}}$, and relative proton momentum $\delta_{\mathrm{HMS}}$, are made to ensure that only particles within the well understood region of the spectrometer momentum acceptance are used.  The momentum in the spectrometers is measured relative to the central momentum $p_{\mathrm{set}}$, so that particles with the same $\delta = (p-p_\mathrm{set})/p_{\mathrm{set}}$ are dispersed by the same amount.

Some parts of the SOS spectrometer acceptance, due to an ambiguity in the solution of the optics equations, do not reconstruct reliable tracks.  The cuts on the SOS focal plane position in the magnet dispersion direction,  $X_{\mathrm{SOS,f.p.}}$, are to eliminate these regions.  The particle identification cuts are described in Section~\ref{pidsec}.  

\begin{table}[!hbtp]
\begin{center}
\begin{tabular}{lrcl}
\hline
\hline
Quantity & Variable & & Cut\\
\hline
Electron momentum		& $\delta_{\mathrm{SOS}}$			& $<$ & $ +20$\% \\
				&					& $>$ & $ -15$\% \\
Proton momentum			& $\delta_{\mathrm{HMS}}$			& $<$ & $ +9$\% \\
				&					& $>$ & $ -9$\% \\
SOS focal plane position	& $X_{\mathrm{SOS,f.p.}}$			& $>$ & $ -20$ cm\\
~~~~dispersive direction	&					& $<$ & $ +22$ cm\\
$^{\dag}$Coincidence time	& $|t_{\mathrm{coin}} - t_{\mathrm{cent}}|$	& $<$ & $ 1.5$~ns\\
$^{\dag}$SOS \v Cerenkov	& $N_{\mathrm{p.e.}}$			& $>$ & $ 0.5$ \\
$^{\dag}$SOS calorimeter	& $E_{\mathrm{norm}}$			& $>$ & $ 0.7$ \\
\hline
\hline
\end{tabular}
\end{center}
\caption[Cuts on data]{\label{tab:cuts}The set of `standard' cuts applied to the data and to the simulations where applicable.  $^{\dag}$The Particle Identification cuts are not applied to the simulation.}
\end{table}

\subsubsection{Binning}
\label{binning}

The data were binned in $W$, cos$\theta^*_{\eta}$, $\phi^*_{\eta}$ and $m_x^2$, where $W$ is the invariant mass of the hadronic system, $\theta^*_{\eta}$ is the polar angle between the direction of the $\eta$ and the three-momentum transfer vector $\vec{q}$ in the centre-of-mass of the resonance, $\phi^*_{\eta}$ is the azimuthal angle of the $\eta$ with respect to the electron scattering plane, and $m_x^2$ is the square of the missing mass for $p(e,e'p)x$.  

For the lower-$Q^2$ data, this was done in 12 cos$\theta^*_{\eta}$-bins and 8 $\phi^*_{\eta}$-bins, to maximise the angular resolution for partial-wave analyses, necessitating $m_x^2$-bins of 0.1~GeV$^2$ and $W$-bins of 30~MeV near the resonance and 40~MeV at higher $W$.  The higher $Q^2$ data, with far fewer detected particles, was binned with $W$-bins of 30 MeV, 6 cos$\theta^*_{\eta}$-bins, 5 $\phi^*_{\eta}$-bins and $m_x^2$-bins of 0.15~GeV$^2$.

Bins in  ($W$, cos$\theta^*_{\eta}$, $\phi^*_{\eta}$) were retained for the analysis if they passed the following three criteria.  Firstly, in the region of the $\eta$ missing mass peak, the simulation was required to predict a signal to background ratio of at least 0.25.  Secondly, the simulation needed to have predicted a minimum average number of $\eta$ events in the peak of 1.5 per missing mass squared channel.  This criterion was used instead of requiring a total number of predicted $\eta$ particles because the resolution of the missing mass peak changes substantially with cos$\theta^*_{\eta}$.  The third criterion for acceptance was, following the subtraction of the all the backgrounds, the sum of the data in the region of the missing mass peak was required to have a statistical uncertainty of less than 50\%.

\subsection{Particle Identification}
\label{pidsec}

\subsubsection{Electron Identification}

\label{sec.sospid}

In the SOS spectrometer, the \v Cerenkov detector and the electromagnetic calorimeter were used to identify electrons and reject pions.  The \v Cerenkov detector was filled with Freon-13 at 1 atmosphere, yielding a velocity threshold of $\beta_{t}=1/n=0.9992$.  The highest momenta detected by the SOS in this experiment was about 2.09~GeV/$c$, corresponding to $\beta=0.9978$ for pions, which is below the threshold for detection while all electrons are well above the threshold.  Some pions make small signals in the \v Cerenkov due to scintillation or ``knock-on" electrons from atomic scattering.  The detected signal was calibrated into units of the number of photo-electrons, $N_{\mathrm{p.e.}}$.

For each event, the signals from each of the 44 lead-glass blocks in the calorimeter were summed to obtain the total energy deposited, $E_{\mathrm{tot}}$.  This energy was then normalised by the momentum of the particle as determined by the tracking, $p_{\mathrm{track}}$, to obtain $E_{\mathrm{norm}} = E_{\mathrm{tot}} / p_{\mathrm{track}}$.  The 16 radiation lengths of lead-glass bring electrons to a stop, resulting in a peak at $E_{\mathrm{norm}} \sim 1$ due to electrons.  The pions peak at about $E_{\mathrm{norm}} \sim 0.25$, but have a long tail to higher $E_{\mathrm{norm}}$ due to the charge exchange nuclear interaction $\pi^-p\rightarrow \pi^0nx$, and subsequent decay $\pi^0\rightarrow\gamma\gamma$.

\begin{figure}[!hbt]
\begin{center}
\includegraphicstwowidths{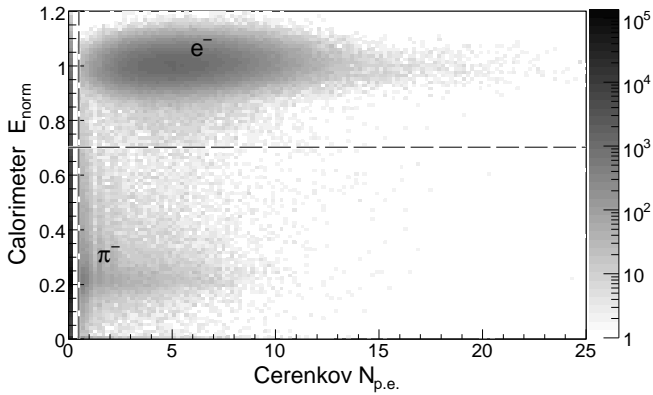}{0.48}{0.96}
\caption[Electron particle identification plot]{The correlation between $E_{\mathrm{norm}}$ and $N_{\mathrm{p.e.}}$ for all the lower-$Q^2$ data.  The particle ID cuts to select electrons, $N_{\mathrm{p.e.}}>0.5$ and $E_{\mathrm{norm}} > 0.7$, are visible as dashed lines in the figure.  All other cuts listed in Table~\ref{tab:cuts} have already been applied to the data.}
\label{pidfig}
\end{center}
\end{figure}

Figure~\ref{pidfig} shows the correlation between $E_{\mathrm{norm}}$ and $N_{\mathrm{p.e.}}$ for the lower-$Q^2$ data.  The electrons are clearly well separated from the pions by these two detectors.  In the analysis, electrons are identified using two simple cuts, $N_{\mathrm{p.e.}}>0.5$ and $E_{\mathrm{norm}} > 0.7$, shown in the figure. 

\subsubsection{Proton Identification and Accidental Coincidence Subtraction}

\begin{figure}[!hbt]
\begin{center}
\includegraphicstwowidths{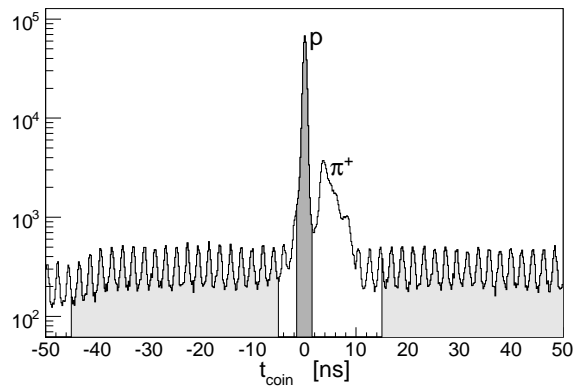}{0.48}{0.96}
\caption[Corrected coincidence time]{Coincidence time spectrum for the lower-$Q^2$ data with all of the `standard' cuts except the coincidence time cut.  The dark grey shaded region represents the 3~ns wide proton cut.  The 2~ns beam structure is clear in the accidental background.  Data from the light grey shaded regions was used to estimate the amount of accidentals in each ($W$, cos$\theta^*_{\eta}$, $\phi^*_{\eta}$ and $m_x^2$) bin under the proton peak.}
\label{cointime}
\end{center}
\end{figure}

Protons were separated from pions using time of flight considerations.  The raw difference in arrival times, $t_{\mathrm{diff}}$, between the electron in the SOS and the positive particle in the HMS, were corrected event-by-event for differences in path length of both particles through the detectors and the variation in velocity $\beta$ of the positive particle (all electrons having essentially the same velocity.)  This corrected coincidence time, $t_{\mathrm{coin}}$, is plotted in Fig.~\ref{cointime} and shows peaks due to protons and $\pi^{+}$ particles and a background of accidental (or random) coincidences.

The path taken is determined by the tracking algorithm from drift chamber hit positions while the velocity $\beta = (p^2/(m_p^2 + p^2))^{1/2}$ is calculated from the measured momentum $p$ assuming the proton mass $m_{p}$.  For protons the corrected coincidence time depends only on the actual difference in starting times of the particles in the target, causing a peak of real coincidences, which has been shifted to zero in the figure.  Particles with a different mass, such as pions, have their coincidence time peak shifted relative to the protons since for the same momentum, they have a different velocity.  The $\pi^{+}$ peak is broader than the proton peak because $t_{\mathrm{coin}}$ is calculated to remove the momentum dependence of the protons but the pion locus remains momentum dependent.  A much smaller number of kaons are detected and form a locus between the pions and protons, but remain distinctly separable.  It was then possible to select the proton events and reject the pion and kaon events and most of the accidental coincidences using one simple cut.

Accidental coincidences occur when both detectors are triggered within the 100~ns coincidence time window, but the detected particles originate in different scattering events.  In the coincidence time spectrum of Fig.~\ref{cointime}, the accidentals are the continuous background under the two main peaks.  The 2~ns beam structure can clearly be seen in the spectrum.  A 3~ns particle identification window was used to select protons, but within this cut there is still some background due to accidental coincidences which must be subtracted.  

For each bin in 4 dimensions ($W$, cos$\theta^*_{\eta}$, $\phi^*_{\eta}$ and m$^2_x$), the number of accidental coincidences inside the proton cut was estimated by determining the average number of accidentals in the ``wings'' of the spectrum, $-45\mathrm{~ns}<t_{\mathrm{coin}}<-5\mathrm{~ns}$ and $15\mathrm{~ns}<t_{\mathrm{coin}}<50 \mathrm{~ns}$, away from loci for actual coincidences.  This value was then normalised for the width of the proton cut and subtracted from the data.  The accidental correction is small for our kinematics and rates, the weighted mean correction was 1.5\% and the largest correction in any ($W$, cos$\theta^*_{\eta}$, $\phi^*_{\eta}$) bin was 7.6\%.

\subsubsection{$\eta$ Identification}
\label{sec.etaID}

In the case of inelastic scattering, the detection of the scattered electron and recoil proton is not an exclusive measurement---there will be at least one other emitted particle.  If there is only one undetected particle it is possible to fully reconstruct the kinematics of that particle. The data corresponding to such a channel, the $p(e,e'p)\eta$ in this case, is isolated by constructing the square of the missing mass $m_x^2$, as given in Eq.~(\ref{eqn.mmass2}).  

\begin{figure}[!hbtp]
\begin{center}
\includegraphicstwowidths{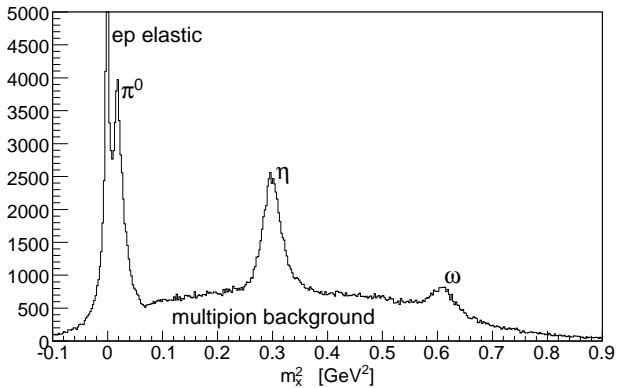}{0.48}{0.96}
\caption[Missing mass squared, lower-$Q^2$]{Missing mass squared $m_x^2$, from Eq.~(\ref{eqn.mmass2}), for all the lower-$Q^2$ data.}
\label{fig.mmass2}
\end{center}
\end{figure}

Figure~\ref{fig.mmass2} shows the $m_x^2$ distribution for the lower-$Q^2$ data, with the $\pi^0$, $\eta$ and $\omega$ products are visible as peaks.  The energy calibration of the data is done on the elastic peak.  A simple Gaussian and polynomial background fitted to the eta peak
is centered at 0.2993 GeV$^2$, within 0.2\% of the nominal eta mass.  The broadening is due to instrumental resolution.

The actual extraction of the $\eta$ particles is done by applying a cut on $m_x^2$ around the $\eta$ peak and subtracting the background.  The resolution of this peak varies as a function of cos$\theta^*_{\eta}$ and therefore so does the cut, which is listed in Table~\ref{tab:mmass2cutslowQ2} for the lower-$Q^2$ data.  The higher-$Q^2$ data has very little coverage above cos$\theta^*_{\eta}$ = 0 at any $W$, and larger $m_x^2$ bins, so the cut was kept at a constant 0.255~GeV$^2 < m_x^2 < 0.36$~GeV$^2$.  

The continuous background, seen in Fig.~\ref{fig.mmass2}, is due to events with more than one undetected particle.  In this case, the missing mass does not correspond to any physical mass because the magnitude of the missing momentum is smaller than the sum of the magnitudes of the individual momenta of the undetected particles.  This effect, predominantly due to the production of multiple pions, is the principle background in this experiment, and is treated in Section~\ref{multipi}.

\begin{table}[!hbtp]
\begin{center}
\begin{tabular}{l|cccccc}
\hline
\hline
\backslashbox{\phantom{1234567}}{cos$\theta^*_{\eta}$} &  -0.917 & -0.750 & -0.583 & -0.417 & -0.250 & -0.083 \\
\hline
$m_x^2$ min (GeV$^2$)		& 0.27 & 0.27 & 0.27 & 0.26 & 0.26 & 0.25 \\
$m_x^2$ max (GeV$^2$)		& 0.34 & 0.35 & 0.35 & 0.36 & 0.36 & 0.37 \\
\hline
\hline
\backslashbox{\phantom{1234567}}{cos$\theta^*_{\eta}$} &  ~0.083 & ~0.250 & ~0.417 & ~0.583 & ~0.750 & ~0.917 \\
\hline
$m_x^2$ min (GeV$^2$)		&  0.25 & 0.25 & 0.25 & 0.25 & 0.25 & 0.25 \\
$m_x^2$ max (GeV$^2$)		&  0.37 & 0.38 & 0.38 & 0.39 & 0.39 & 0.39 \\
\hline
\hline
\end{tabular}
\end{center}
\caption[Missing mass squared cuts vs. cos$\theta^*_{\eta}$, lower-$Q^2$]{\label{tab:mmass2cutslowQ2}  The $m_x^2$ cuts used in each cos$\theta^*_{\eta}$ bin for the lower-$Q^2$ data.}
\end{table}

\subsection{Monte Carlo Simulation of the Experiment}
\label{sec.montecarlo}

The Monte Carlo simulation of the experiment was done with SIMC~\cite{absimc}, the Jefferson Lab Hall C in-house detector simulation package.  The simulation includes detailed models of both magnetic spectrometers and simulated the effects of radiative processes, multiple scattering, and ionisation energy loss (due to material in the target and spectrometers).  It was used to obtain the experimental acceptance and radiative corrections for the resonance process under study, to simulate the multipion background to the resonance production and to study a number of other processes serving to verify our understanding of the apparatus.

SIMC as a package consists of an event generator, which is able to produce events from a variety of physical scattering processes common  in Hall C or from phase space, and two `single arm' spectrometer models, one for each detector, to track the particles and determine whether they are accepted by the detector.  In each spectrometer model, the particle is propagated from its initial position in its initial direction with transport maps produced by COSY Infinity~\cite{Makino:2006sx}, an arbitrary-order, beam dynamics simulation and analysis code, using the results of a field map of the magnetic elements.  At points where there are apertures in the spectrometer such as collimators or the magnets themselves, the positions of the particles are checked against these.  For the magnets this is done at the entrance, exit and at the maximum beam envelope within the object.  Particles making it into the detector hut underwent multiple scattering and energy loss in the air and other materials.  Particles that didn't conform to the experimental trigger, such as passing through three of the scintillator hodoscopes, and for electrons the \v Cerenkov and calorimeter, were considered undetected.  Detected events were reconstructed back to the target using the COSY optics matrix.

SIMC was not used `out of the box' for the present analysis, as it didn't have physics models for either the p(e,e'p)$\eta$ process or for multiple pion production.  For $\eta$ production, a simple model of the $S_{11}$ resonance was added to SIMC, which was then run to simulate the signal part of the experiment.   In the case of the multipions, another event generator was used and the resulting electron and proton pairs were propagated through the SIMC detector models to simulate their detection.  Both the elastic $ep$ and the $\pi^{0}$ peaks have radiative tails that extend into the region $m_x^2 > 0.1 $ GeV$^2$.  A Monte Carlo simulation determined that this makes a negligible contribution to the $\eta$ peak compared to the multipion background.

For both of the two $Q^2$ configurations, the data are taken in ``settings'' for which the HMS spectrometer angle and momentum is fixed. To limit file sizes and aid in online checking of the data, the data in each setting is taken in a number of ``runs''.  The simulation is performed on a run-by-run basis to match the data.  The data and simulation are then binned into identical four-dimensional histograms.

\subsubsection{Model for p(e,e'p)$\eta$}
\label{sec.sigmodel}

The model for $\eta$ production used in the simulation and extraction of the cross-section is a single relativistic Breit-Wigner shape as a function of $W$ multiplied by a exponential form factor depending on $Q^2$.  The form used for the Breit-Wigner resonance shape (from Christy and Bosted~\cite{Christy:2007ve}) is given by 
\begin{eqnarray}
\mathrm{BW}(W) & = & \frac {K_R K^{cm}_R}{ K(W) K^{cm}(W)} \cdot \nonumber\\
& & ~~\cdot\frac{\Gamma^{\rm tot} \Gamma^{\gamma} }{ \Gamma
\left [ (W^2 - W_R^2)^2 + (W_R \Gamma^{\rm tot})^2 \right ]}, 
\end{eqnarray}
where the equivalent photon energy in the lab frame is \[K(W) = \frac{(W^2 - m_p^2) }{  2 m_p},\] the equivalent photon energy in the center of mass (CM) frame is \[K^{cm}(W) = \frac{(W^2 - m_p^2) }{ 2 W},\] and $K_R$ and $K^{cm}_R$ represent the same quantities evaluated at the mass of the $S_{11}$ resonance, $W_R$.  $\Gamma^{\rm tot}$ is the full decay width defined by 
\begin{equation}
\Gamma^{\rm tot} = \sum_j \beta_j \Gamma_j,
\end{equation}
with $\beta_j$ the branching fraction to the $\rm j^{th}$ decay mode and $\Gamma_j$ the partial width for this decay mode. 
The partial widths are determined from the intrinsic widths $\Gamma$, using
\begin{equation}
 \Gamma_j = \Gamma \left [\frac  { p^{cm}_j }{ p^{cm}_j |_{W_R}} 
 \right ]^{2L+1} \cdot 
\left [ \frac  { (p^{cm}_j|_{W_R})^2 +  X^2}{ (p^{cm}_j)^2  + X^2} \right ]^L,
\end{equation}
where $p^{cm}_j$ is the momentum, in the center of mass, of a meson produced by a system of invariant mass $W$ and $p^{cm}_j|_{W_R}$ is the momentum of a meson from a system of the nominal resonance invariant mass $W_R$,  $L$ is the angular momentum of the resonance, and $X$ = 0.165 GeV is an empirical damping parameter.  The model as used in the simulation is then given by 
\begin{equation}
\label{eqn:bwQ2}
\frac{d\sigma}{d\Omega^*_\eta} = \frac{1}{4\pi}a e^{-bQ^2}\cdot \mathrm{BW}(W).
\end{equation}

Although simplistic, the model describes the data well.  The parameters $a$ and $b$ were obtained by fitting the form $a e^{-bQ^2}$ to the cross-section at the $S_{11}$ resonance mass, $\sigma_{R}$, of data taken by Armstrong \emph{et al.}~\cite{armstrong99} and both of the present $Q^2$ data sets.  The parameters $W_{R}$ and $\Gamma_{R}$ were refined using an iterative procedure in which the Breit-Wigner form was fitted to the angle-integrated lower-$Q^2$ data, used to extract a new cross-section and then refitted.   There was no explicit cos$\theta^*_{\eta}$ or $\phi^*_{\eta}$ dependence in the input model since the data showed very little anisotropy.  The final model parameters are given in Table~\ref{tab.param}.

\begin{table}[!hbtp]
\begin{center}
\begin{tabular}{cc}
\hline
\hline
Parameter	& Value\\
\hline
$a$ 		& $9.02$ nb \\
$b$		& $-0.479$ (GeV/$c$)$^{-2}$\\
$W_{R}$		& $1525$ MeV\\
$\Gamma_{R}$	& $133$ MeV\\
$X$		& 0.165 GeV\\
\hline
\hline
\end{tabular}
\end{center}
\caption[$\eta$ production model parameters]{\label{tab.param}The parameters of the $S_{11}$ resonance-dominated cross section model used for the final data extraction.}
\end{table}

\subsubsection{Model for Multipion Production}

The multipion background was simulated using an event generator from the Jefferson Lab Hall B (CLAS detector) simulation package, which takes as input the $Q^2$ and $W^2$ ranges of the generation region and the reactions, chosen from a list of possibilities, from which the events should be generated.  Depending on the reaction, the events are then sampled from interpolated data tables or according to a cross-section model---in contrast to SIMC behaviour which throws events uniformly and weights them event-by-event.  The generator itself extrapolates the cross-section from where data exists to higher $Q^2$ using the square of the dipole form, $(1 + Q^2/0.71)^{-4}$.  The reactions included in our simulation of the multipion background are given by Eqs.~(\ref{eqnmultipi1}) and (\ref{eqnmultipi2}).

\begin{eqnarray}
\label{eqnmultipi1}
e + p  \rightarrow &  e' + p + \pi^{+}\pi^{-}\phantom{\pi^{+}\pi^{-}\pi^{0}}	&\mathrm{(model)}\\
\nonumber\\
\label{eqnmultipi2}
e + p  \rightarrow &  e' + p + \pi^{+}\pi^{-}\pi^{0}\phantom{\pi^{+}\pi^{-}}		&\nonumber\\
e + p  \rightarrow &  e' + p + \pi^{+}\pi^{-}\pi^{+}\pi^{-} \phantom{\pi^{0}}	&\mathrm{(tables)}\\
e + p  \rightarrow &  e' + p + \pi^{+}\pi^{-}\pi^{+}\pi^{-}\pi^{0}		&\nonumber
\end{eqnarray}

The event generator was developed from an initial version for real photons~\cite{Corvisiero:1994wz}.  In that version, for performance reasons, the cross-section is drawn from tabulated data---either measured or generated from models in unmeasured regions.  In the  current version, the $p\pi^{+}\pi^{-}$ exit channel, Eq.~(\ref{eqnmultipi1}), is now generated according to a phenomenological model~\cite{Mokeev:2001qg}, with parameters that have been fit to recent CLAS data~\cite{PhysRevLett.91.022002} which measured the process $ep \rightarrow e'p\pi^{+}\pi^{-}$ for $1.4 < W < 2.1$~GeV and $0.5 < Q^2 < 1.5$~GeV$^2$/$c^2$.  The model is calculated for the three intermediate channels $\pi^{-}\Delta^{++}$, $\pi^{+}\Delta^{0}$ and $\rho p$.  The amplitude is defined in the meson-baryon degrees of freedom, and is therefore not necessarily valid at this high momentum transfer $Q^2 \lesssim $ 7 (GeV/$c$)$^2$, where quark-gluon degrees of freedom may be the most appropriate.  Radiative corrections are not implemented for the multipion model.  Despite these last two points, the results obtained are good enough to justify our implementation here.  The properties of the generated pions aren't used, just the electron and proton pairs are propagated through SIMC.  

\subsubsection{Multipion Background Subtraction}
\label{multipi}

As can be seen in Fig.~\ref{fig.mmass2}, the peak at $m_x^2\sim0.3$~GeV$^2$, corresponding to missing $\eta$ particles, lies on a continuous background described in Sec.~\ref{sec.etaID}.  This background was treated by simulating the $m_x^2$ spectra of the background using SIMC with a model of the largest contributing reactions, described in the previous section, and then subtracting the simulation from the data.  

\begin{figure}[!hbtp]
\begin{center}
\includegraphicstwowidths{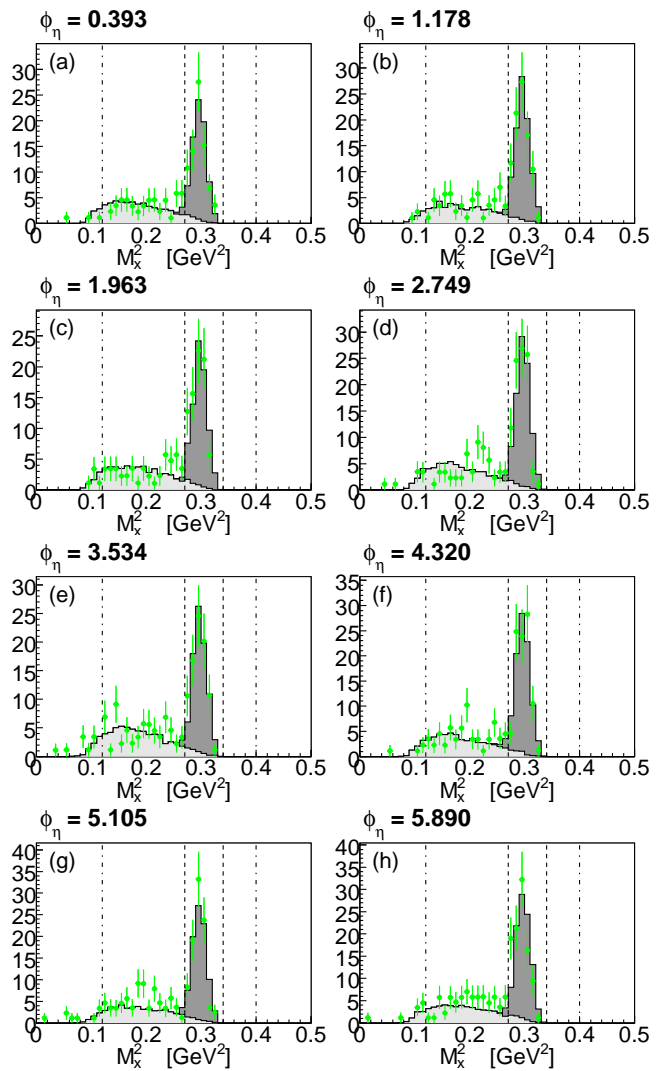}{0.48}{0.63}
\caption[$\phi_{\eta}$ dependence of missing mass squared distributions]{(Colour online) The $\phi_{\eta}$ dependence of missing mass squared distributions for $W=1.5$ GeV and cos$\theta^*_{\eta}=-0.916$.  The (green) points are the data while the solid fill is the sum of the simulations.  The multipion background component is the light grey filled area and $\eta$ production simulation component has the darker grey fill.  The dot-dashed lines shows the region within which the background fit is done while the dashed lines show the region within which the $\eta$ cross-section is extracted}
\label{fig.mmass2phidep}
\end{center}
\end{figure}

The output of the simulation was a large set of multipion events that are accepted into our detectors.  These events are then filled into histograms of the same structure as those of the data, yielding our approximation to the shape of the multipion background, without an absolute normalisation.  Since an absolute multipion cross-section is not being extracted, the shape is sufficient to subtract it from the data.  It can be seen in Fig.~\ref{fig.mmass2phidep}, showing the data and associated simulations for one ($W$, cos$\theta^*_{\eta}$) bin, that excellent agreement is obtained.

\begin{figure}[!hbtp]
\begin{center}
\includegraphicstwowidths{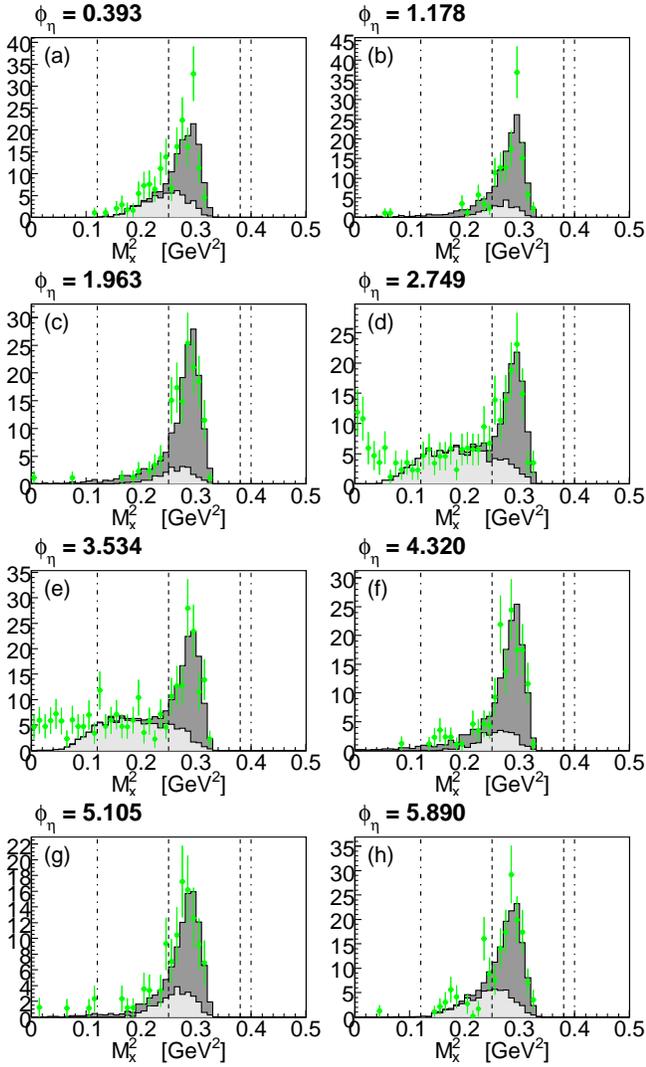}{0.48}{0.63}
\caption[$\phi_{\eta}$ dependence of missing mass squared distributions]{(Colour online) The $\phi_{\eta}$ dependence of missing mass squared distributions for $W=1.5$ GeV and cos$\theta^*_{\eta}=0.416$.  Symbols as in Fig.~\ref{fig.mmass2path}.  Panels with $\phi_{\eta} =$ 1.178, 1.963, 4.320, and 5.105 are the out-of-plane $\phi$ bins where the simulations of the signal and background are sufficiently similar to make a two-parameter bin-by-bin fit unreliable.}
\label{fig.mmass2path}
\end{center}
\end{figure}

The simplest way to normalise the background to the data is with a two-parameter fit in each ($W$, cos$\theta^*_{\eta}$, $\phi^*_{\eta}$) bin.  The $m_x^2$ spectra of the multipion background simulation and the $\eta$ production simulation would have been normalised to minimise the $\chi^2$ difference between their sum and data $m_x^2$ spectrum.  In practice, due to diminishing acceptance, the out-of-plane $\phi^*_{\eta}$ bins demonstrate a phenomenon where the multipion background simulation and the $\eta$ production simulation can have $m_x^2$ spectra similar enough to make a two-parameter fit unreliable.  This is typically the case for mid to large cos$\theta^*_{\eta}$ and worsens as $W$ increases.  An example of such a case is illustrated in Fig.~\ref{fig.mmass2path}.  

For this reason, the fit was constrained to have the multipion normalisation parameter constant over $\phi^*_{\eta}$, as expected physically.  For each and all of the ($W$, cos$\theta^*_{\eta}$) bins, the fit had 9 parameters: one for the single multipion normalisation over all the $\phi^*_{\eta}$ bins and one for $\eta$ production in each of the eight $\phi^*_{\eta}$ bins.  The production of $\pi^0$ particles, seen as a peak at $m_x^2\sim 0.02$ GeV$^2$ in some panels of Figs.~\ref{fig.mmass2path} and \ref{fig.mmass2thetadep}, produces a radiative tail which, in principle, extends under the $\eta$ peak.  The size of this effect is smaller than the uncertainty in the multipion background, and so was neglected.  

\ifthenelse{\equal{\twocol}{false}}{
	\begin{sidewaysfigure}[!hbtp]
}{
	\begin{figure*}[!hbtp]
}
\begin{center}
\includegraphics[width=0.999\textwidth]{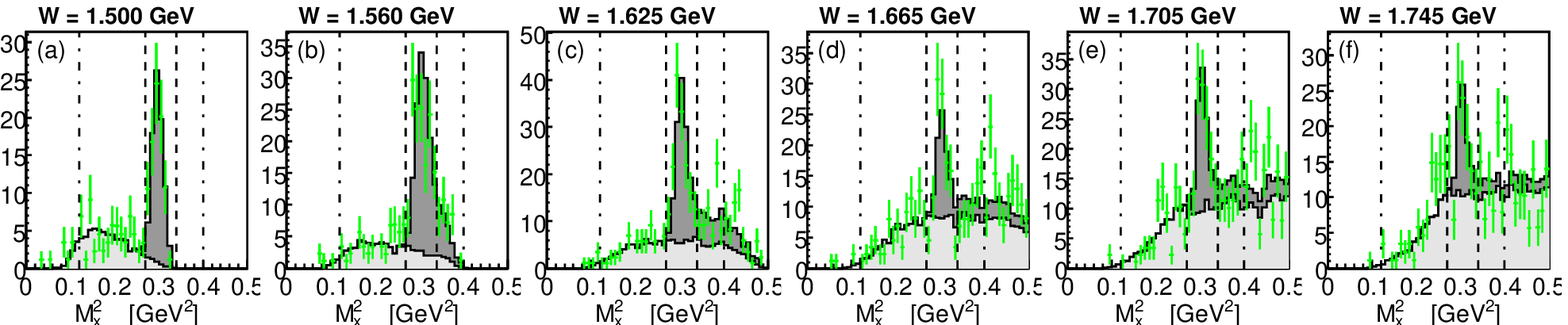}
\caption[$W$ dependence of missing mass squared distributions]{(Colour online) The $W$ dependence of missing mass squared distributions for cos$\theta^*_{\eta}=-0.916$ and $\phi=3.534$~radians.  Symbols as in Fig.~\ref{fig.mmass2path}.}
\label{fig.mmass2Wdep}
\end{center}
\ifthenelse{\equal{\twocol}{false}}{
	\end{sidewaysfigure}
}{
	\end{figure*}
}

\ifthenelse{\equal{\twocol}{false}}{
	\begin{sidewaysfigure}[!hbtp]
}{
	\begin{figure*}[!hbtp]
}
\begin{center}
\includegraphics[width=0.999\textwidth]{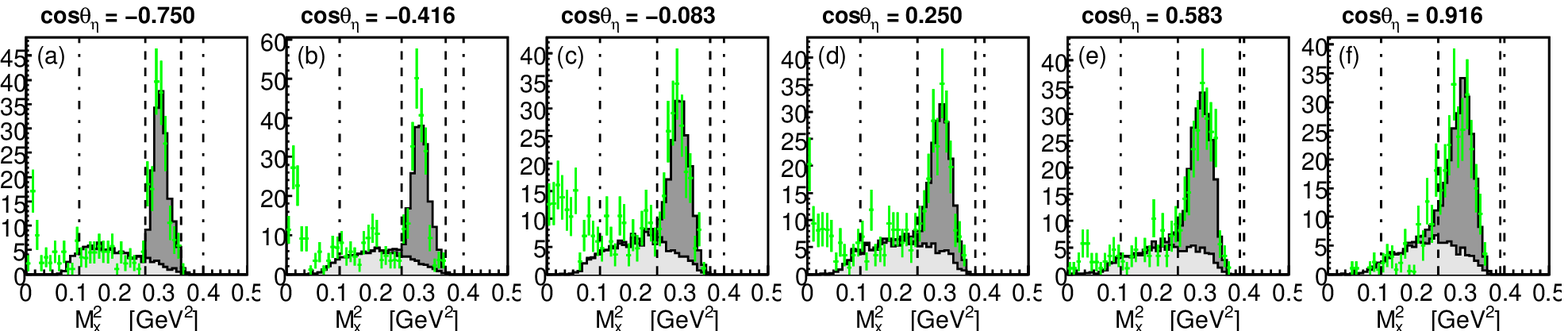}
\caption[cos$\theta^*_{\eta}$ dependence of missing mass squared distributions]{(Colour online) The cos$\theta^*_{\eta}$ dependence of missing mass squared distributions for $W=1.5$ GeV and $\phi=3.534$~radians.  Symbols as in Fig.~\ref{fig.mmass2path}.  The vertical dashed lines, which show the region within which the $\eta$ cross-section is extracted, vary with cos$\theta^*_{\eta}$ to accommodate the changing resolution.}
\label{fig.mmass2thetadep}
\end{center}
\ifthenelse{\equal{\twocol}{false}}{
	\end{sidewaysfigure}
}{
	\end{figure*}
}

This approach does a good job of reproducing the shape of the measured $m_x^2$ spectra.  By eye, the sum of the normalised simulations seem to match the data well and in 94\% of bins have a reduced $\chi^2$ of less than 2.  A few representative spectra showing the $W$ and cos$\theta^*_{\eta}$ dependence of the $m_x^2$ distributions are shown in Figs.~\ref{fig.mmass2Wdep} and~\ref{fig.mmass2thetadep} respectively.  The uncertainty in the normalised background simulation was determined by adding the small Monte Carlo statistical uncertainty to the Minuit~\cite{James:1975dr} fit uncertainty on the normalisation parameter in quadrature.

\begin{figure}[!hbtp]
\begin{center}
\includegraphicstwowidths{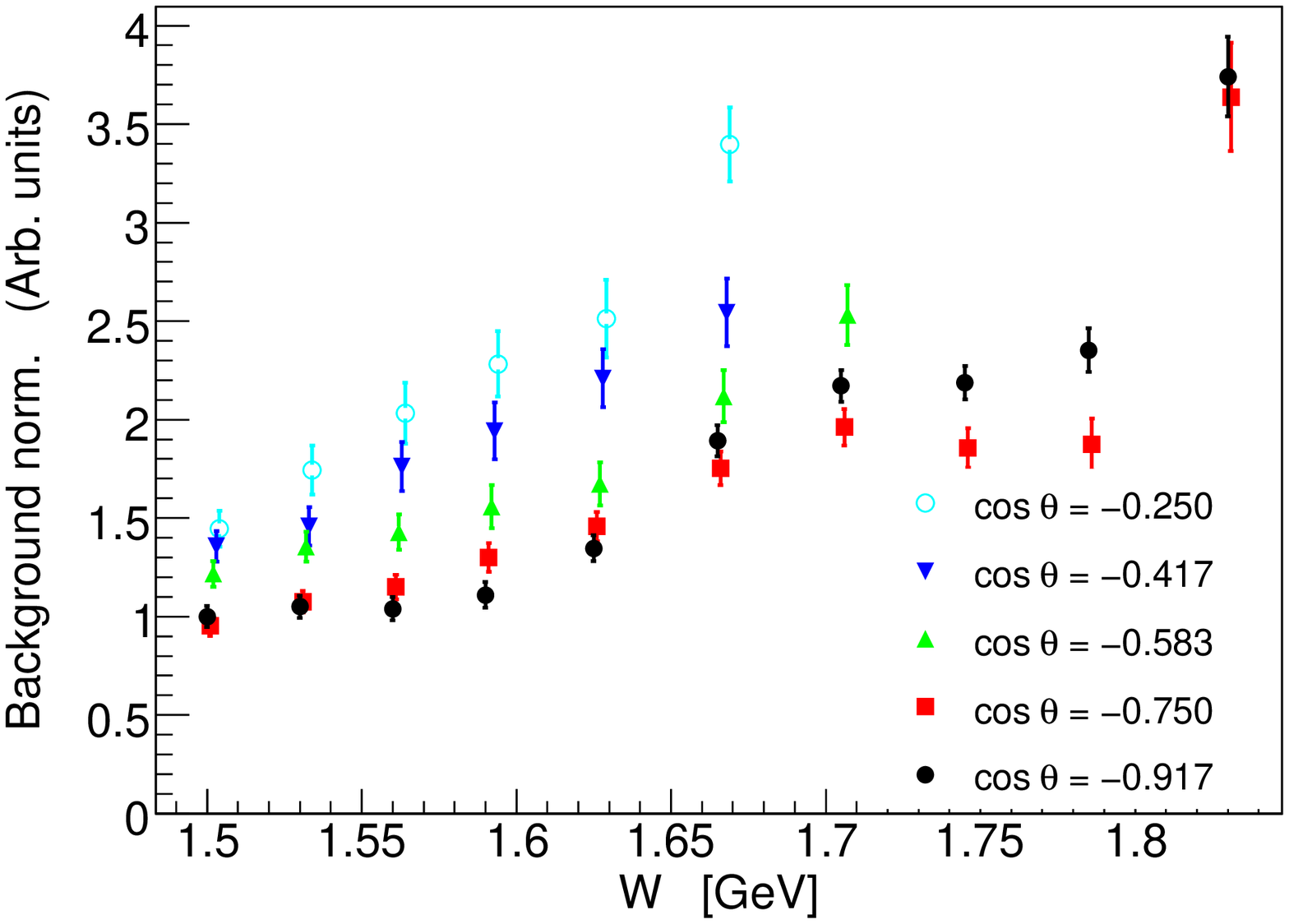}{0.48}{0.8}
\caption[$W$ dependence of multipion background normalisation]{(Colour online) The $W$ dependence of the normalisation coefficient of the multipion background simulation.}
\label{fig.thetmpi}
\end{center}
\end{figure}

\begin{figure}[!hbtp]
\begin{center}
\includegraphicstwowidths{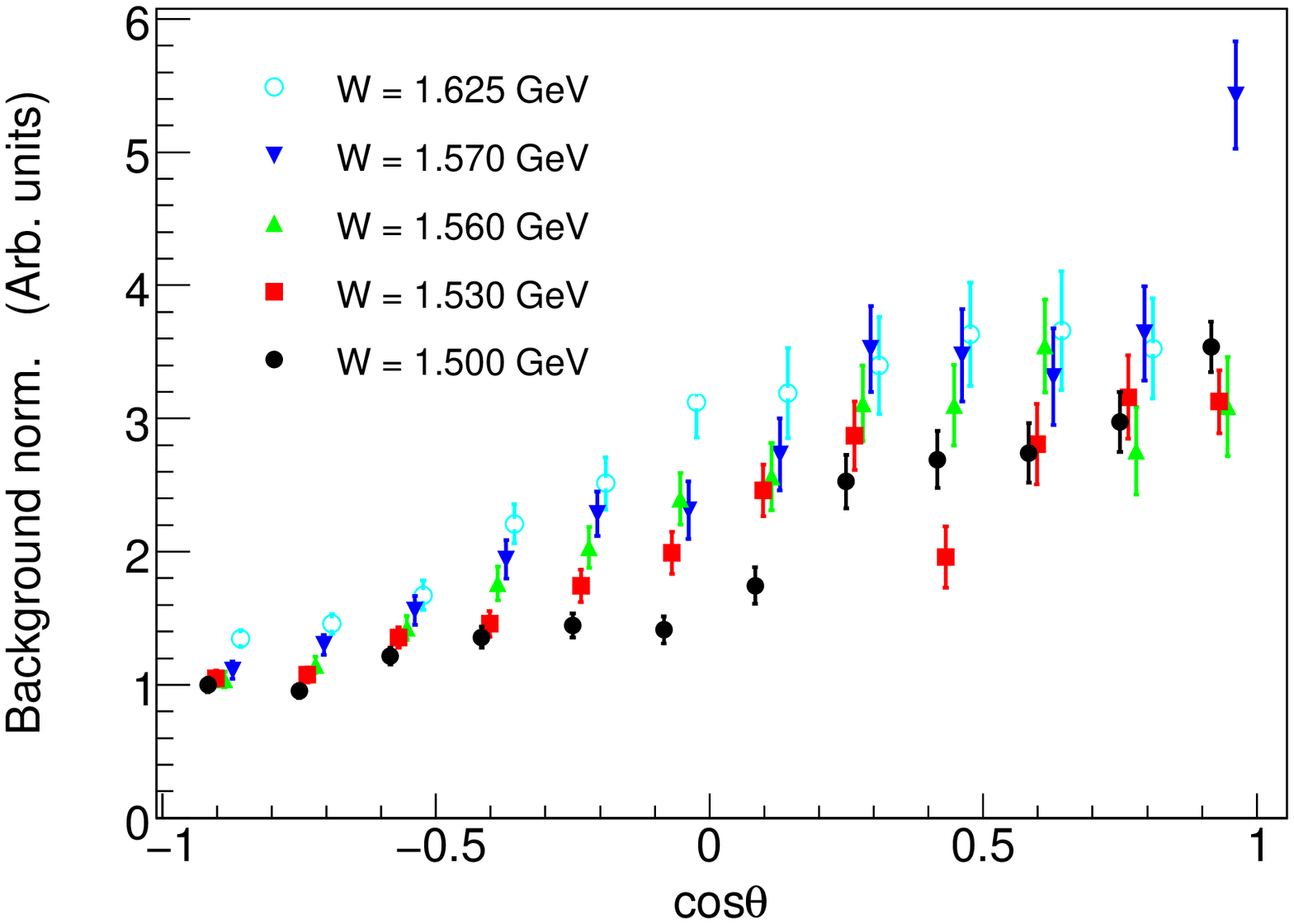}{0.48}{0.8}
\caption[cos$\theta^*$ dependence of multipion background normalisation]{(Colour online) The cos$\theta^*$ dependence of the normalisation coefficient of the multipion background simulation.}
\label{fig.Wmpi}
\end{center}
\end{figure}

It should be noted that some structure is seen within the  normalisation parameters of the background model in $W$ and cos$\theta^*$, illustrated in Figs.~\ref{fig.thetmpi} and~\ref{fig.Wmpi} respectively.  The extracted fit parameters seem to rise smoothly and approximately linearly with both increasing $W$ and increasing cos$\theta^*$.  This is understandable since the multipion background model is produced from data with much lower $Q^2$.  Overall, the variation in the parameters is about a factor of 4.

\subsubsection{Target Window Background}

No explicit subtraction for scattering off the aluminium walls of the target was performed.  The data taken with the dummy target in this experiment has too low statistics to be used for subtraction, and it was not taken at all of the experimental settings, but it is adequate for estimating the yield from the target walls and demonstrating the shape of the missing mass distribution.

The size of the target wall effect is small and it has a very similar shape to the multipion background, so it is therefore adequately accounted for in that background subtraction procedure.  To first order, a nucleus is a bag of nucleons, and as such the multipion production from the aluminium target window has the same broad kinematic distribution as from a free proton---the following analysis confirms this.

\begin{figure}[!hbtp]
\begin{center}
\includegraphicstwowidths{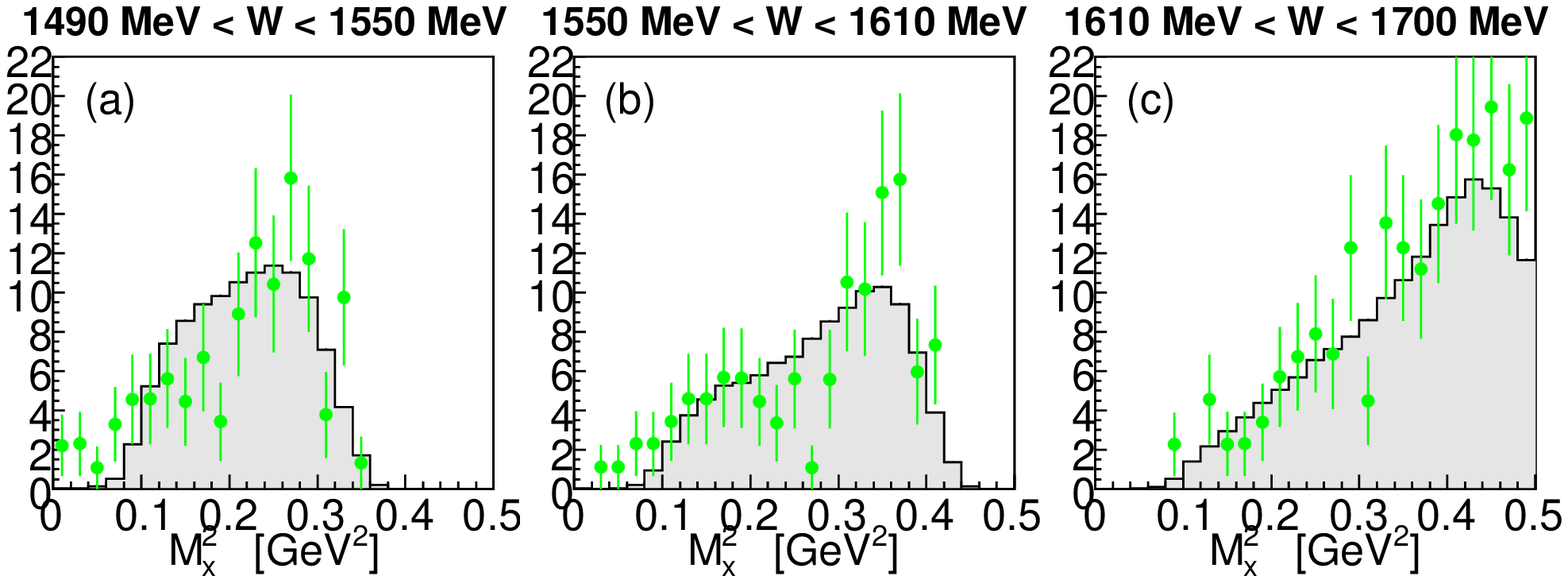}{0.48}{0.96}
\caption[`Dummy' data and multipion background simulation]{(Colour online) The (green) points are the $m_x^2$ distribution of the full set of data off the ``dummy'' target cell, and the grey filled histogram is the simulation of multipion background from Hydrogen.  The simulation is arbitrarily normalised to match the data, with the same factor in all three panels.  Note the similarity in shape.}
\label{fig.dummy}
\end{center}
\end{figure}

The dummy target produced 430 coincidences from a beam charge of 1.97 C giving an average yield, integrated over all angles and $W$ up to 1.7~GeV, of about 0.2 counts per mC.  The hydrogen target's 64,000 multipion coincidences, estimated from the background subtraction procedure, came at about 0.6 counts per mC, or three times as fast.  Taking into account the differences in thickness between the dummy and the actual target walls, the multipion background is expected to have produced at least 20 times more background events than the target walls.  Figure~\ref{fig.dummy} shows the similarity between missing mass spectra of the dummy data and the multipion background simulation in three W bins. 

\subsubsection{Radiative Corrections}
\label{sec.radcorr}

Radiative effects occur because photons are emitted in the interaction of the incoming and outgoing charged particles of the scattering.  These real photons are either produced within the field of the scattering nucleus itself, called internal radiation, or from the fields of other nuclei in the propagation medium, called external radiation.  This radiation causes there to be a difference between the actual momenta of the particles at the scattering vertex and the detected momenta, leading to measured values of $W$, $Q^2$ and the c.m. angles cos$\theta^*_{\eta}$ and $\phi$, different from that of the actual scattering.  In order to extract meaningful information from the detected particles, this radiative contribution must be corrected for.  

External radiation is small for the proton due to its high mass and can be handled essentially exactly for the electron, both pre and post-scattering.  Dealing with internal radiation requires a knowledge of the coupling of the photon to the electron, which is well known, and to the proton, which isn't known analytically since it depends on its QCD structure.  It is then further complicated by interference of the amplitudes for radiation from each of the particles of the scattering.  The radiative corrections for this experiment are done within SIMC, with the formalism of Ref.~\cite{Ent:2001hm}, which is a general framework for applying radiative corrections in $(e,e'p)$ coincidence reactions at GeV energies.  This approach uses the angle peaking approximation and takes into account higher-order bremsstrahlung effects, multiple soft photon emission and radiation from the scattered hadron.  External radiation is also included in the model.  

The size of the radiative corrections implemented by SIMC is determined by running the full simulation with and without including radiative effects.  In each bin, the ratio of the number of events predicted by these two simulations, after the `standard' cuts of Table~\ref{tab:cuts} and the missing mass cuts of Sec.~\ref{sec.etaID}, gives a number equivalent to the correction factor required to take account of the radiative effects.  This radiative correction factor is listed for each bin in Tables~\ref{tab.cslowQ2} and ~\ref{tab.cshighQ2} along with the extracted cross-sections.  Using these values and the size of the missing mass cuts given in Sec.~\ref{sec.etaID} one can remove the effect of the radiative corrections on the cross-sections.

\begin{figure}[!hbtp]
\begin{center}
\includegraphicstwowidths{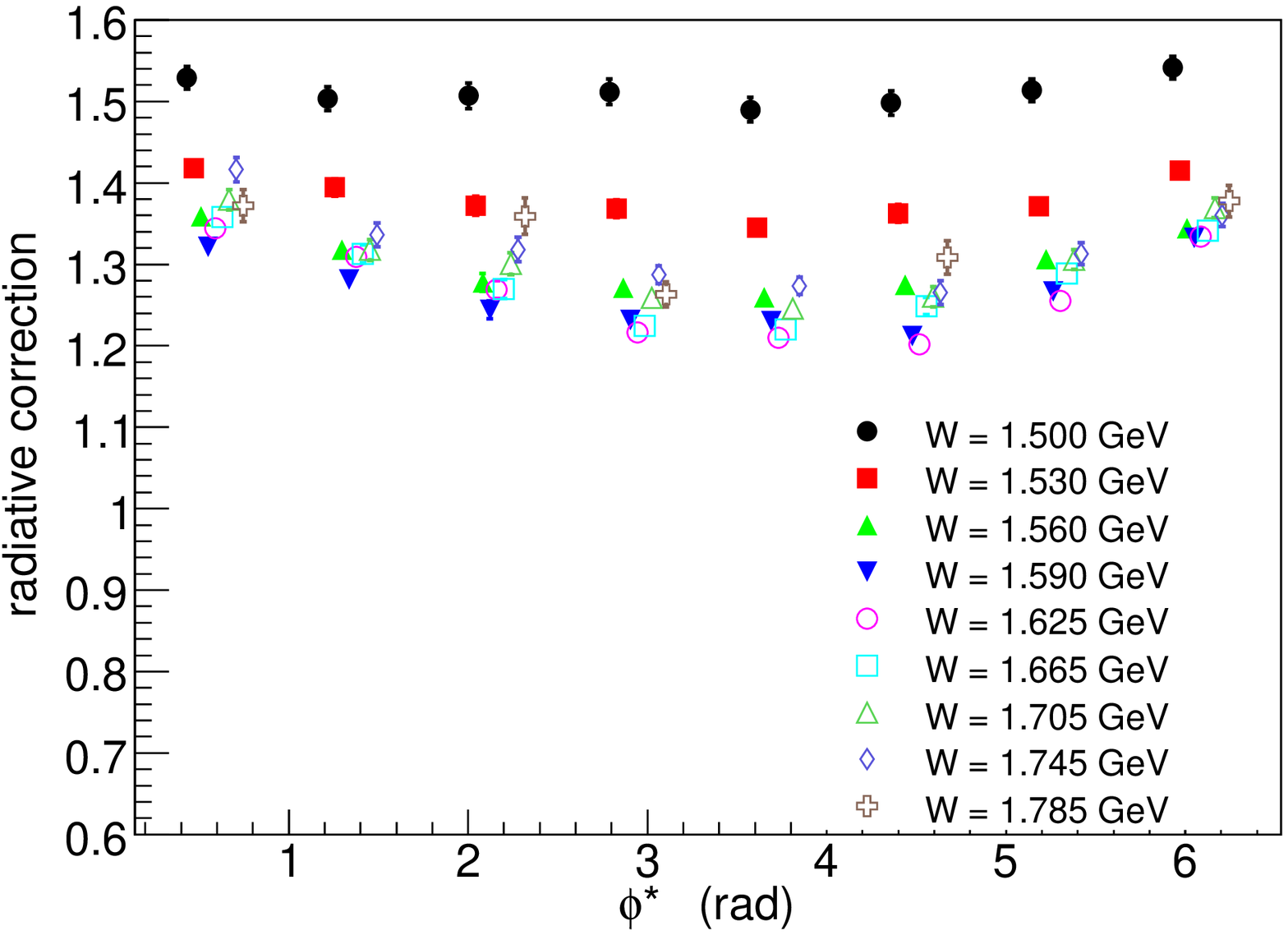}{0.48}{0.8}
\caption[Radiative corrections for -1 $<$ cos$\theta_{\eta}$ $<$ -$\frac{1}{3}$]{(Colour online) Radiative corrections for -1 $<$ cos$\theta_{\eta}$ $<$ -$\frac{1}{3}$.  Uncertainty is due to Monte Carlo statistics only.}
\label{fig.radcorr1}
\end{center}
\end{figure}

\begin{figure}[!hbtp]
\begin{center}
\includegraphicstwowidths{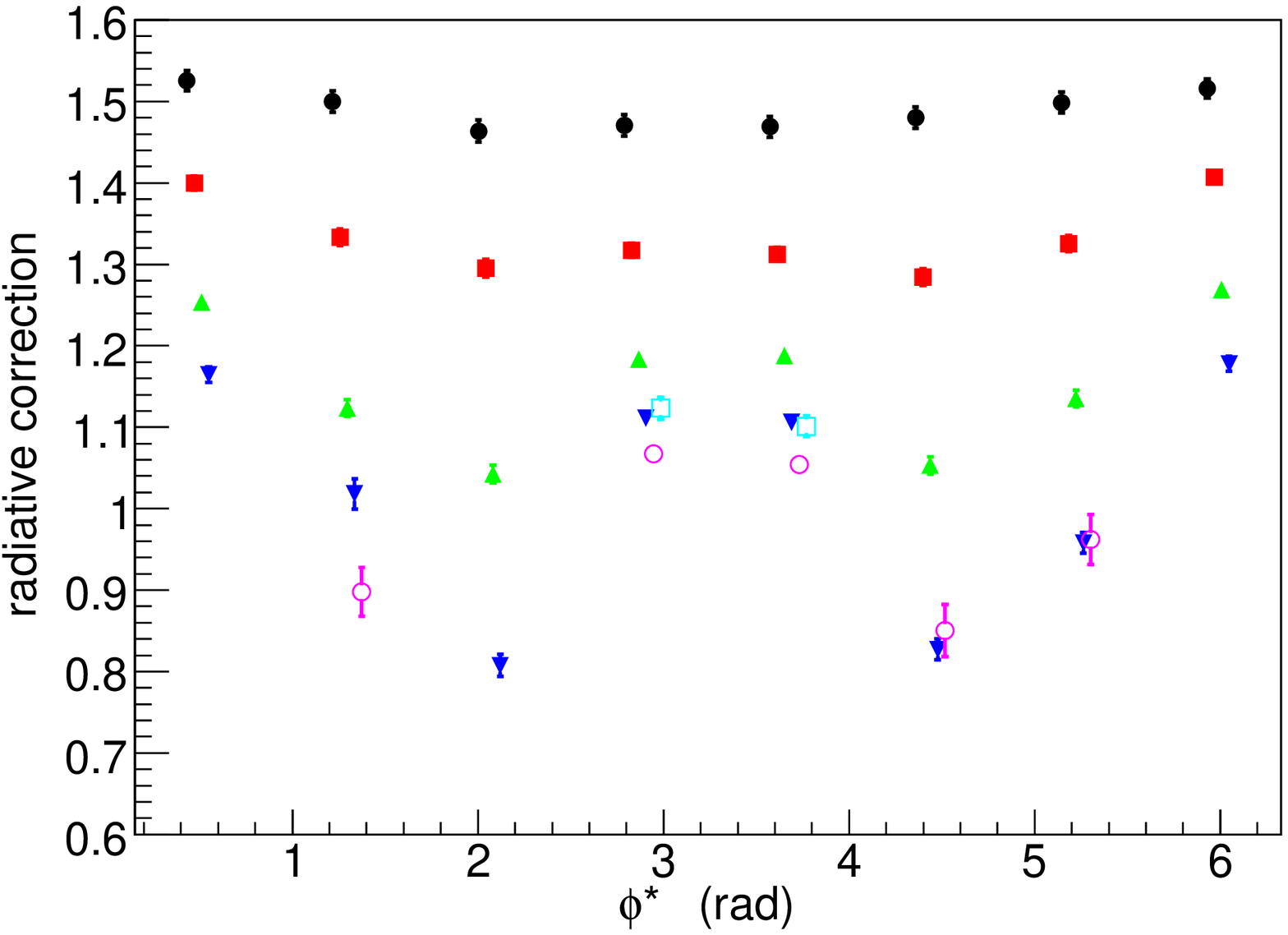}{0.48}{0.8}
\caption[Radiative corrections for -$\frac{1}{3}$ $<$ cos$\theta_{\eta}$ $<$ $\frac{1}{3}$]{(Colour online) Radiative corrections for -$\frac{1}{3}$ $<$ cos$\theta_{\eta}$ $<$ $\frac{1}{3}$.  Uncertainty is due to Monte Carlo statistics only.  Symbols as in Fig.~\ref{fig.radcorr1}.}
\label{fig.radcorr2}
\end{center}
\end{figure}

\begin{figure}[!hbtp]
\begin{center}
\includegraphicstwowidths{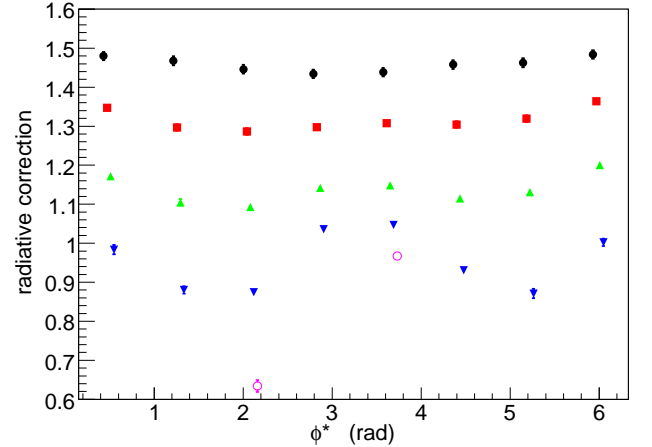}{0.48}{0.8}
\caption[Radiative corrections for $\frac{1}{3}$ $<$ cos$\theta_{\eta}$ $<$ 1]{(Colour online) Radiative corrections for $\frac{1}{3}$ $<$ cos$\theta_{\eta}$ $<$ 1.  Uncertainty is due to Monte Carlo statistics only.  Symbols as in Fig.~\ref{fig.radcorr1}.}
\label{fig.radcorr3}
\end{center}
\end{figure}

The correction factor is plotted for the lower-$Q^2$ configuration as a function of $\phi$ for different $W$ bins and three cos$\theta_{\eta}^*$ ranges in Figs.~\ref{fig.radcorr1}, \ref{fig.radcorr2} and \ref{fig.radcorr3}.  The points are plotted for the kinematic bins where the data are sufficient to extract a cross section.  Much of the large kinematic dependence in these plots comes about due to the limited acceptance, which decreases with increasing $W$ and cos$\theta_{\eta}^*$.  

This approach does neglect 2-photon radiation, which is expected to be about a factor of $\alpha=1/137$ smaller, and makes approximations.  The uncertainty in the radiative corrections was estimated to be 2\%.

\subsection{Extraction of the $\eta$ Differential Cross-Section}

The actual $\eta$ cross-section extraction is done by comparing the data, having had the randoms and multipion background already subtracted, with a Monte Carlo simulation of the experiment, produced using SIMC and the $\eta$ production model described previously.  The comparison is done for each ($W$, cos$\theta^*_{\eta}$, $\phi^*_{\eta}$) bin.  The $m_x^2$ dependence of both the subtracted data and the simulation is integrated out between two tight limits in $m_x^2$ that contain the missing $\eta$ particle peak
\[
\label{integral}
N^i = \sum_{j_{\mathrm{low}}}^{j_{\mathrm{high}}}N^{ij},
\]
where $i$ labels the $(W, \mathrm{cos}\theta^*_{\eta}, \phi^*_{\eta})$ bins and $j$ labels the $m_x^2$ bins so that $N^{ij}$ is the content of a certain $(W, \mathrm{cos}\theta^*_{\eta}, \phi^*_{\eta}, m_x^2)$ bin.  The number of simulation events, $N^i_{\mathrm{MC}}$, is obtained by multiplying the yield output of SIMC, in counts per mC, by the integrated beam current and then using the same filling procedure.

As can be seen in Fig.~\ref{fig.mmass2thetadep}, the resolution of the experiment, and therefore the width of the $\eta$ peak, depends on cos$\theta^*_{\eta}$.  The integration limits, $j_{\mathrm{low}}$ and $j_{\mathrm{high}}$---the dashed lines in the figure, are also functions of cos$\theta^*_{\eta}$.  The dependence of the extracted cross-sections on these integration limits is accounted for in the next section.  The experimental cross-section is then obtained, from the model cross-section at the centre of the bin $\sigma^i_{\mathrm{MC}}$, using

\begin{equation}
\label{datamc}
\sigma^i_{\mathrm{data}}=\frac{N^i_{\mathrm{data}}}{N^i_{\mathrm{MC}}}\sigma^i_{\mathrm{MC}}.
\end{equation}

As with any measurement in which the events are histogrammed, each bin represents a mean quantity, weighted by the distribution of the events within that bin.  In this experiment the cross-section changes rapidly and non-linearly with $W$, especially going from threshold to maximum within just 50 MeV, and our $W$ bins are rather large at 30 MeV.

The bin centering in $W$ was done implicitly during the cross-section extraction, under the assumption that the relativistic Breit-Wigner model and SIMC are accurate representations of the physics and detector response. 
 If the simulation experimental acceptance model is accurate, then the kinematic distribution of simulated particles in each bin will mimic the population of data events within that bin.  So too if the physics model is good, then nonlinearities in the actual cross-section will be correctly reproduced by the simulation.  To the extent that both of these are true, the ratio of the data and Monte Carlo yields in each bin, $N^i_{\mathrm{data}}/N^i_{\mathrm{MC}}$, directly connects the number of detected particles with the input Monte Carlo model, $\sigma^i_{\mathrm{MC}}$.  The bin centring is then done by evaluating the simulation input model at the bin centre.  

The bins in cos$\theta^*_{\eta}$ and $\phi^*_{\eta}$ are quite small,  and where there is full coverage, the extracted differential cross-sections are largely flat.  It was decided not to attempt to incorporate nonlinear variation of the angular cross-section into the input model, and thus no implicit bin centering takes place. 

The results are not quoted at fixed $Q^2$.  Since the events in every bin have a $Q^2$ distribution, the cross-section results are an average over the $Q^2$ distribution of the bin.  The weighted average $Q^2$ of events in each bin $\langle Q^2_{\mathrm{bin}}\rangle$, is therefore quoted along with the extracted cross-section in Tables~\ref{tab.cslowQ2} and \ref{tab.cshighQ2}.  In order to quote all the data at a single value of $Q^2$, a model dependent correction would have to be applied to the data, which can be done at a later stage.

\subsection{Check of SOS acceptance}

\subsubsection{Coincident Elastic Scattering Cross-Section}
\label{sec.elastic}

For  the SOS central momentum and angle
setting of $\theta_{SOS} = 47.5^{\circ}$
and $P_{SOS} = 1.74$~GeV/c, the scattered protons
from elastic $ep$ events will have a momentum of 4.44~GeV/c
and angle of $18.3^{\circ}$. The elastic electrons
cover an electron momentum range of 2.08 to 1.73~GeV/c
and angular range of $44^{\circ}$ to $51^{\circ}$
which corresponds to a proton momentum range of
4.25 to 4.61~GeV/c and angular range of 19.8$^{\circ}$
to 17.0$^{\circ}$. The $Q^2$ range is from 6.4 to 7.1~(GeV/c)$^2$.

During the experiment, the HMS was set at  three
combinations of $\theta_{HMS}$  and  $P_{HMS}$ at
which elastic $ep$ coincidence events were detected.
At $\theta_{HMS} = 18^{\circ}$ and $P_{HMS} = 4.7$~GeV/c,
the acceptance for elastic $ep$ events is best matched.
At $\theta_{HMS} =  19.5^{\circ}$ and $P_{HMS} =4.5$~GeV/c,
the HMS in-plane angular acceptance reduces the SOS in-plane angular
range to $44^{\circ}$ to $47.5^{\circ}$. While
for   $\theta_{HMS} =  16.5^{\circ}$ and $P_{HMS} =4.5$~GeV/c,
the HMS in-plane angular acceptance reduces the SOS in-plane angular
range to $49^{\circ}$ to $51^{\circ}$.

To extract measured elastic $ep$ yields,
the same data cuts listed in Table~\ref{tab:cuts} were used with an additional
 cut of $ 0.8 < W < 1.07$~GeV to isolate elastic events. The
data were also corrected for tracking efficiency, trigger inefficiency, computer and electronic deadtime.
The same SIMC Monte Carlo was used with $ep$ elastic cross section
calculated using the electric and magnetic form factors from
the fit of Bosted~\cite{Bosted:1994tm}.  At this $Q^2$ = 6.76~(GeV/c)$^2$, the
proton magnetic form factor is the dominant contribution to the
elastic cross section and a conservative estimated
error on the predicted cross section is 4\%.

\begin{figure}[!hbt]
\begin{center}
\includegraphics[angle=270, width=0.48\textwidth]{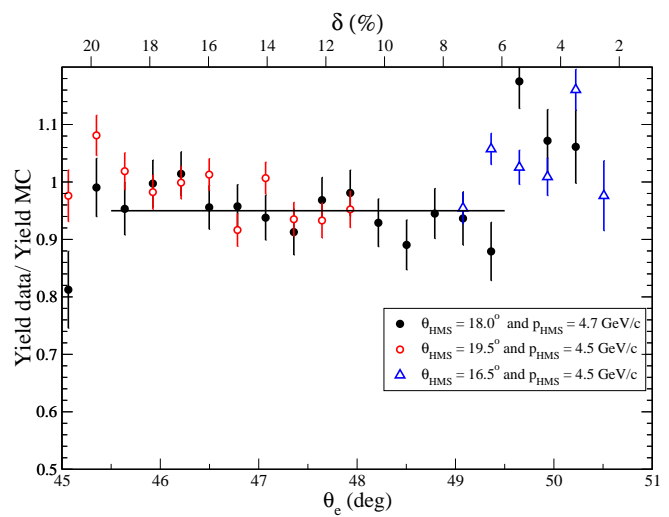}
\caption[Elastic $ep$ coincidence vs. $\theta_{e}$ (lower-$Q^2$)]{(Colour online) Ratio of yield of elastic $ep$ coincidence events to predicted yield from Monte Carlo (Yield Data/Yield MC) plotted versus $\theta_{e}$ for $\theta_{SOS}$ = 47.5$^{\circ}$ and three different combinations  of $\theta_{HMS}$ and $p_{HMS}$. The solid line is the average ratio = 0.95  $\pm$ 0.01, of all points between $\theta_{e}$ = 45.5$^{\circ}$ to 49.5$^{\circ}$. The corresponding value of electron $\delta$ for a given $\theta_{e}$ is given by the upper x-axis.}
\label{fig.ep}
\end{center}
\end{figure}

In Fig.~\ref{fig.ep}, the ratio of
data yield to predicted Monte Carlo yield is plotted as a function of
electron scattering angle for all three settings. Between  scattered
electron angle of $45.5^{\circ}$ to $49.5^{\circ}$, the
ratio is reasonably constant with an average value of 0.95 $\pm$ 0.01
which indicate good agreement with previous measurements.
Below $45.5^{\circ}$, the agreement falls off sharply and
above $49.5^{\circ}$ the ratio jumps to an average of 1.08 which
demonstrate problems in understanding the SOS acceptance in some areas.

On the other hand, Figure~\ref{fig.ep} shows that we are able to reproduce a well known quantity, the elastic scattering cross section, to within a few percent using our two spectrometer coincidence configuration, and thus we develop some confidence in the main result of the paper.  The elastic events are in the SOS relative momentum range $10\%<\delta<20\%$, while the $p(e,ep)\eta$ cross section is extracted in the range $-20\%<\delta<-5\%$, thus we cannot use this data to correct the $\eta$ cross section.  For this reason a single arm comparison is best for checking the SOS acceptance. The $ep$ coincidence comparison is useful as a check on the understanding of the experimental luminosity and efficiency corrections.

\subsubsection{Inclusive Elastic and Inelastic Cross-Section}
\label{sec.inclusive}

In order to determine how accurately the SIMC simulation package models the acceptance of the SOS spectrometer, we extracted single-arm elastic and inelastic cross-sections from hydrogen and compared them with a fit to previous data.  This inclusive analysis had the same set of data runs, the same correction factors whenever applicable, the same acceptance simulation code and the same electron identification cuts, as the coincidence analysis.

In the inclusive case, corrections for the target endcaps were much larger than in the coincidence case, and an additional correction for pair-symmetric backgrounds was needed (up to 10\% at the highest $W$).   These were determined using interpolated positron cross sections measured in a previous experiment~\cite{Malacephd} with the same target and beam energy, but slightly different scattering angles at 45, 55, and 70 degrees.  This correction is negligible for the coincidence analysis due to the imposition of missing mass cuts. 

Another difference is that radiative corrections were done analytically, rather than in the Monte Carlo simulation.  For both elastic and inelastic scattering these were calculated using the formalism of Mo and Tsai~\cite{Mo:1968cg}. For the required elastic scattering cross-section model, we used the form factor parametrisation of Bosted~\cite{Bosted:1994tm}, while for the inelastic cross-section model we used the May 2007 fit of Christy and Bosted~\cite{Christy:2007ve}.

To obtain final radiated cross-section for a proton target, the cross-sections from the Al dummy target were subtracted with the appropriate scale factor to match the thickness of the endcaps.  The small difference in radiative corrections between the endcaps and dummy was not taken into account.

\begin{figure}[!hbt]
\begin{center}
\includegraphicstwowidths{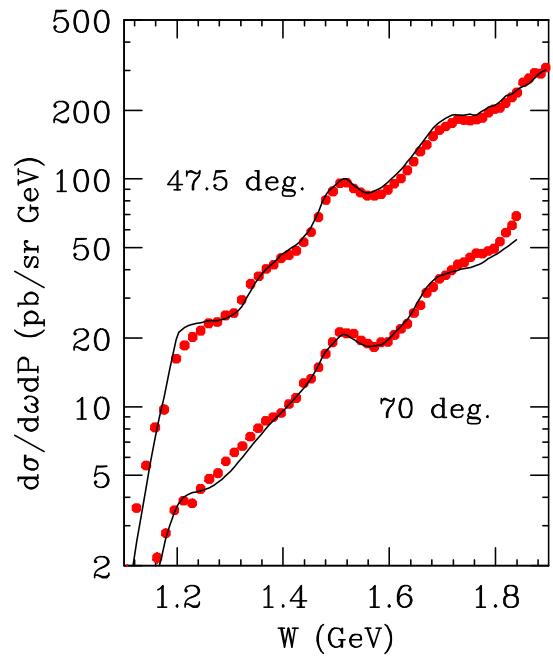}{0.4}{0.8}
\caption[Inclusive inelastic differential cross-sections]{(Colour online)
Inclusive inelastic differential cross-sections as measured
by the SOS spectrometer centered at 47.5 and 70 degrees,
as a function of $W$, with the angular cut $-30<dy/dz<30$~mr.
The curves are from a fit to world data~\protect{\cite{Christy:2007ve}}.}
\label{fig.sossig}
\end{center}
\end{figure}

The $W$ dependence of the extracted inelastic cross-section, taken from the central region of the SOS spectrometer is plotted in Fig.~\ref{fig.sossig} along with the Christy model.  
Generally, the $W$-dependence is in quite good agreement with Christy fit, especially for $1.5<W<1.6$ GeV, which is the main focus of this paper.  Based on this analysis, a systematic uncertainty of 3\% was assigned to the acceptance of the SOS spectrometer.

\subsection{Systematic Error Analysis}
\label{sec.errors}

Depending on the source of error, one of two different methods was used to account for it.  Those errors that were independent of the kinematic variables of the extracted data, $W$, cos$\theta^*_{\eta}$ and $\phi^*_{\eta}$, were treated globally and applied to the data overall.  The sources of this kind of error are summarised in Table~\ref{tab.errall}.

The pion contamination through the PID cut for electrons was calculated by Villano~\cite{Villanophd}, using the data of this experiment, to be 1.6\%.  Most of these pions are from random coincidences and are effectively removed by the coincidence time cut---an analysis for the $F_{\pi}$ experiment~\cite{Hornphd,Blok:2008jy} shows the residual contamination to be about 0.1\%.  The systemstic error in the target density and charge measurement were also determined by the $F_{\pi}$ analysis~\cite{Hornphd,Blok:2008jy}.  The error in the HMS acceptance is the quadrature sum of the 0.5\% point-to-point error and 0.8\% normalisation error determined by Christy~\cite{Christy:2004rc}. The overall error of 4.2\%, calculated as a sum in quadrature, is dominated by the uncertainty in the SOS acceptance.

\begin{table}[!hbtp]
\begin{center}
\begin{tabular}{lcl}
\hline
\hline
Parameter			& Uncertainty 	& Reference \\
\hline
SOS acceptance			& ~3.0\%~ 	& Sec.~\ref{sec.inclusive} \\
Radiative Corrections		& 2.0\%		& Sec.~\ref{sec.radcorr}\\
Trigger efficiency		& 1.4\% 	& Sec.~\ref{sec.trigeff} \\
Proton absorption		& 1.0\% 	& Sec.~\ref{sec.trigeff} \\
HMS acceptance			& 1.0\% 	& Ref.~\cite{Christy:2004rc}  \\
Target density			& 0.6\% 	& Ref.~\cite{Blok:2008jy}  \\
Charge measurement		& 0.5\% 	& Ref.~\cite{Blok:2008jy}  \\
Electron PID cut		& 0.1\% 	& Ref.~\cite{Villanophd,Blok:2008jy}  \\
\hline
TOTAL (quadrature sum)		& 4.2\% \\
\hline
\hline
\end{tabular}
\end{center}
\caption[Global systematic errors]{\label{tab.errall}The sources of global systematic error and their estimated sizes.}
\end{table}

If a source of error was expected to be dependent on kinematics, then it was treated on a bin-by-bin basis.  The Monte Carlo simulation was run with altered parameters to mimic the uncertainty, and the subsequent analysis was done to compare to the extracted cross-section and quantify the effect bin-by-bin.  The parameters that were altered, listed in Table~\ref{tab.errbin}, were those considered imprecisely known or that affect the determination of the cross-section.   The best choice set of parameters were used for the standard analysis from which the final differential cross-section was calculated.  Each parameter was then  varied and the complete analysis repeated, up to the point of attaining the differential cross-section.  The parameters were not varied together, as would be done in a fit, since it was assumed that to first order they acted independently and thus the prohibitive extra effort was unnecessary.

The drift chamber resolutions, $r_{\mathrm{DC}}$, for the HMS and SOS spectrometers were calibrated as described in Section~\ref{sec.res}.  In order to completely account for any error, these parameters were arbitrarily increased by 10\% for the variation procedure.  The exact position of the target in the beam direction, $z_{\mathrm{targ}}$, was only known to within 3 mm.  For the standard analysis, the middle position of this uncertainty window, an offset of  1.5~mm from the nominal centre, was chosen.  The variation used for this parameter was the maximum possible extent of the motion, 1.5 mm in either direction.  

The SOS spectrometer was found to be somewhat out-of-plane, but the exact amount is uncertain.  A survey of the hall produced a value of $x'_{\mathrm{SOS}} = 2.62$~mr, which was used in this extraction, while an analysis of $ep$ coincidence data by the $F_{\pi}$ experiment~\cite{Hornphd,Blok:2008jy} yielded $x'_{\mathrm{SOS}} = 3.2$~mr.  The spectrometer offset was thus varied in both directions, to 1.5~mr and 3.5~mr, for the systematic analysis.   

The cut on missing mass squared $m_x^2$, is described in Sec.~\ref{sec.etaID}.  The effect of this cut was taken into account by including it as one of the parameters varied in the systematic analysis.  The variation chosen was to widen this cut on both ends by 0.1~GeV$^2$ and then subsequently to narrow it by the same amount. 

If $x_i$ was the value of the differential cross-section in bin $i$ for the standard analysis and $y_i^v$ was for the analysis of a certain variation $v$, then the systematic error for that variation in that bin was taken as half the difference, $\delta_i^v = |x_i - y_i^v|/2$.  

For the purposes of conveying the size of each of the systematic errors in Table~\ref{tab.errbin}, a measure of the average size $\langle\delta^v\rangle$ is used.  This is the mean systematic error for all bins, weighted by the statistical error of the measurement in each bin 
\[
\langle\delta^v\rangle = \frac{\sum_i\delta_i^v/\sigma_i^2}{\sum_i 1/\sigma_i^2}\mathrm{,}
\]
where $\sigma_i$ is the statistical error of the differential cross-section in bin $i$.

\begin{table}[!hbtp]
\begin{center}
\begin{tabular}{lrcccc}
\hline
\hline
Parameter			& 
						& $p_{\mathrm{std}}$ 	& $p_{\mathrm{var}}$	& $\langle\delta^v\rangle$ \\ 
\hline
HMS $r_{\mathrm{DC}}$		& (mm)		& 0.57 			& 0.66 		& 3.1\% \\
SOS $r_{\mathrm{DC}}$		& (mm)		& 0.35 			& 0.39 		& 3.7\% \\
$x'_{\mathrm{SOS}}$ offset	& (mr)		& 2.62			& 1.5		& 3.1\% \\
				&		& 2.62			& 3.5		& 2.8\% \\
$z_{\mathrm{targ}}$ offset	& (mm)		& 1.5 			& 0.0 		& 3.0\% \\
				&		& 1.5			& 3.0		& 2.9\% \\
$m_x^2$ cut			& (GeV$^2$)	& $f(\mathrm{cos}\theta^*)$	& $f^{\mathrm{max}+0.1}_{\mathrm{min}-0.1}$	& 3.5\% \\
				&		& $f(\mathrm{cos}\theta^*)$	& $f^{\mathrm{max}-0.1}_{\mathrm{min}+0.1}$	& 2.8\% \\
\hline
\hline
\end{tabular}
\end{center}
\caption[Kinematic dependent systematic errors]{\label{tab.errbin}The various sources of kinematic dependent systematic errors considered in the analysis, the standard simulation values $p_{\mathrm{std}}$, the systematic variation $p_{\mathrm{var}}$, and the weighted mean systematic error for all bins, $\langle\delta^v\rangle$.}
\end{table}

 The total bin $i$ systematic error, $\delta^{\mathrm{tot}}_i$,  was determined by adding in quadrature the systematic error for each variation, $\delta^v_i$, and the global systematic errors, $\delta_{\mathrm{glo}}$, to give $\delta^{\mathrm{tot}}_i = \sqrt{\sum_v (\delta^v_i)^2 + \sum\delta_{\mathrm{glo}}^2}$.

\section{Results}
\label{sec.results}

\subsection{Differential Cross-Section $p(e,e'p)\eta$}
\label{sec.results.diff}

\begin{figure*}[!hbtp]
\begin{center}
\includegraphicstwowidths{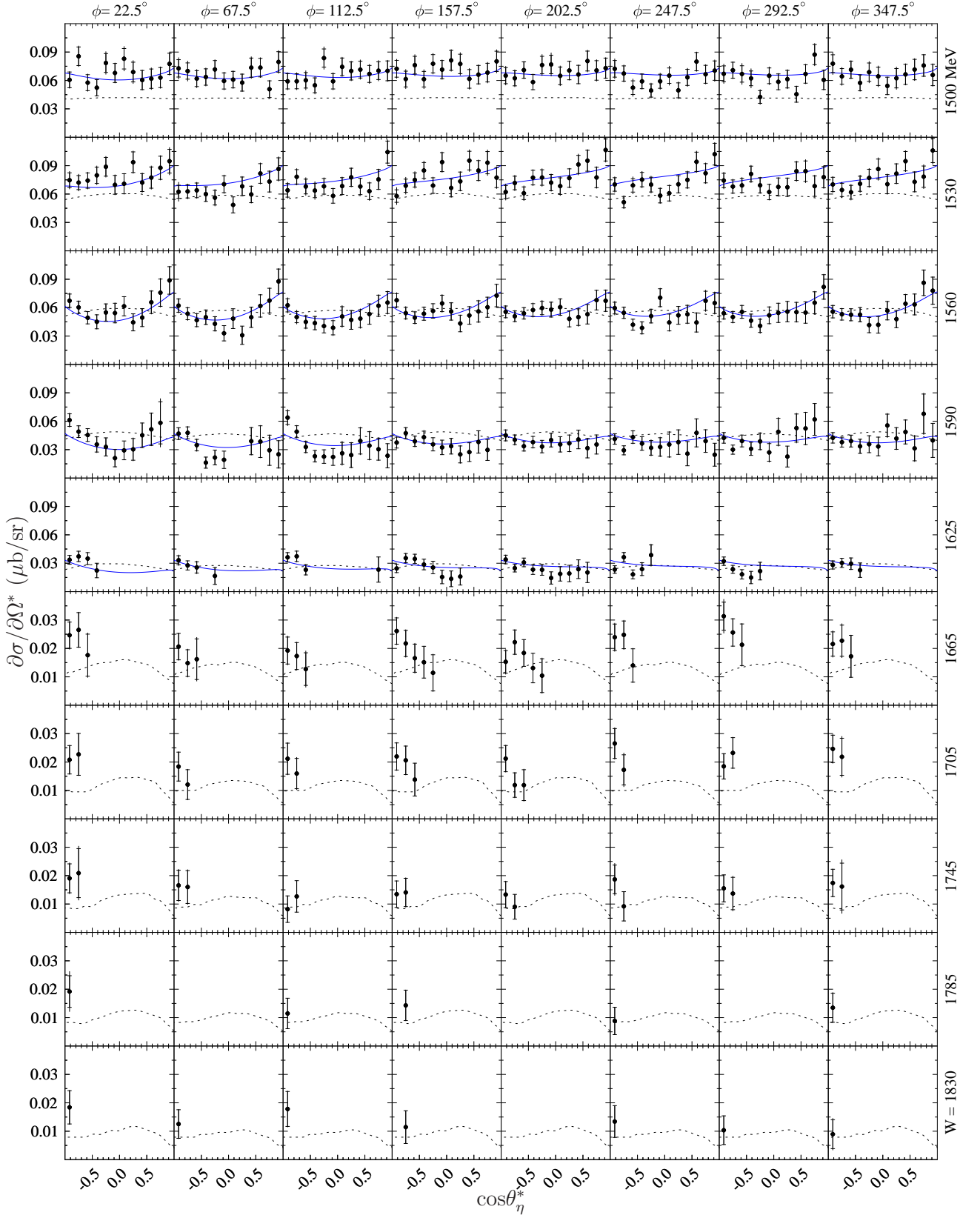}{0.96}{0.96}
\caption[Extracted $ep\rightarrow ep\eta$ differential cross-sections (lower-$Q^2$)]{(Colour online) Extracted $ep\rightarrow ep\eta$ differential cross-sections for the lower-$Q^2$ setting.  The solid (blue) curve is a fit of Eq.~(\ref{eqn.multipole2}) to each $W$ bin.  The dashed curve is the \textsc{eta-maid}~\cite{Chiang:2001as} isobar model for $\eta$-electroproduction from the nucleon at $Q^2$ = 5 GeV$^2$, projected to the appropriate $Q^2$ for each $W$ bin by the factor (5 GeV$^2/Q^2(W))^3$.  The inner error bars are statistical and the outer error bars, the quadrature sum of the statistical and systematic errors.}
\label{fig.lowQ2stamps}
\end{center}
\end{figure*}

The differential cross-sections for the centre-of-mass scattering angles of the $\eta$ are extracted in the bins described in Section~\ref{binning}, with large $W$ bins to allow more angular bins.  Figure~\ref{fig.lowQ2stamps} shows these data for the lower-$Q^2$ setting.  The diminishing experimental acceptance as $W$ increases, especially in out-of-plane $\phi^*_{\eta}$ bins, is evident.  As seen in previous data~\cite{armstrong99,krusche95,brasse84,Denizli:2007tq}, a dominant isotropic, or $S$-wave, component is seen at $W$ from threshold to the $S_{11}$ resonance peak.

\begin{figure}[!hbt]
\begin{center}
\includegraphicstwowidths{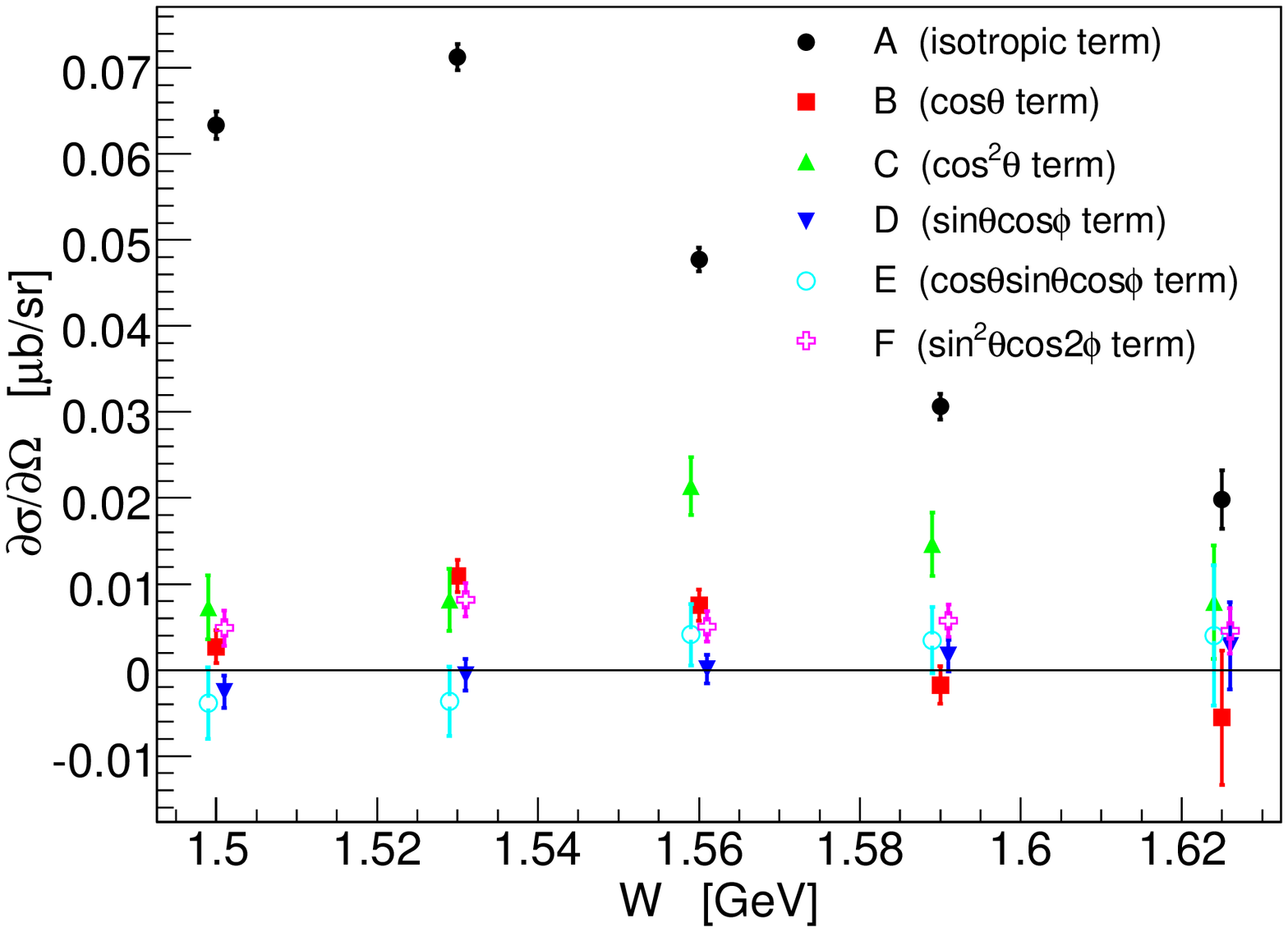}{0.48}{0.96}
\caption[Parameters of angular fit to $\frac{d\sigma}{d\Omega^*_\eta}$ (lower-$Q^2$)]{(Colour online) Extracted parameters from fits of Eq.~(\ref{eqn.multipole2}) to the lower-$Q^2$ differential cross-section, shown as curves in Fig.~\ref{fig.lowQ2stamps}.}
\label{fig.fitpar}
\end{center}
\end{figure}

Equation~(\ref{eqn.multipole2}) is the parametrisation of the virtual photon cross-section in terms of its angular dependence.  The extracted differential cross-section was fitted with Eq.~(\ref{eqn.multipole2}), for the lower $W$ bins where there is sufficient angular acceptance for a fit, and is plotted in Fig.~\ref{fig.lowQ2stamps}.  The parameters extracted from the fit are plotted in Fig.~\ref{fig.fitpar} and listed in Table~\ref{tab.fitpar}.  Using the results of the fit, the anisotropy in the threshold to resonance region is shown to be at most about 15\% for the lower-$Q^2$ setting.

\ifthenelse{\equal{\twocol}{false}}{
	\begin{sidewaystable}[!hbtp]
}{
	\begin{table*}[!hbtp]
}
\begin{center}
\begin{tabular}{l|crclcrclcrclcrclcrcl}
\hline
\hline
	& \multicolumn{4}{c}{W = 1500 MeV}	& \multicolumn{4}{c}{W = 1530 MeV}	& \multicolumn{4}{c}{W = 1560 MeV}	& \multicolumn{4}{c}{W = 1590 MeV}	& \multicolumn{4}{c}{W = 1625 MeV} \\
\hline
A &~& 63.34 & $\pm$ & 1.59 &~& 71.25 & $\pm$ & 1.53 &~& 47.75 & $\pm$ & 1.41 &~& 30.65 & $\pm$ & 1.50 &~& 19.84 & $\pm$ & 3.42\\
B &~& 2.70 & $\pm$ & 1.89 &~& 10.95 & $\pm$ & 1.88 &~& 7.54 & $\pm$ & 1.79 &~& -1.74 & $\pm$ & 2.18 &~& -5.53 & $\pm$ & 7.81\\
C &~& 7.27 & $\pm$ & 3.71 &~& 8.17 & $\pm$ & 3.62 &~& 21.37 & $\pm$ & 3.37 &~& 14.63 & $\pm$ & 3.66 &~& 7.88 & $\pm$ & 6.58\\
D &~& -2.53 & $\pm$ & 1.89 &~& -0.57 & $\pm$ & 1.83 &~& 0.10 & $\pm$ & 1.66 &~& 1.67 & $\pm$ & 1.83 &~& 2.80 & $\pm$ & 5.06\\
E &~& -3.83 & $\pm$ & 4.16 &~& -3.67 & $\pm$ & 4.04 &~& 4.11 & $\pm$ & 3.59 &~& 3.47 & $\pm$ & 3.83 &~& 4.01 & $\pm$ & 8.16\\
F &~& 4.86 & $\pm$ & 2.04 &~& 8.17 & $\pm$ & 1.96 &~& 5.06 & $\pm$ & 1.78 &~& 5.73 & $\pm$ & 1.86 &~& 4.53 & $\pm$ & 2.65\\
\hline
\hline
\end{tabular}
\end{center}
\caption[Angular parameters from fit to lower-$Q^2$ data]{\label{tab.fitpar}The extracted angular parameters from a fit of Eq.~(\ref{eqn.multipole2}) to the lower-$Q^2$ extracted differential cross-section.}
\ifthenelse{\equal{\twocol}{false}}{
	\end{sidewaystable}
}{
	\end{table*}
}

The results of this fit can be compared to similar studies of the angular dependence of $\eta$ production data.  The recent CLAS data~\cite{Denizli:2007tq} was also fit with Eq.~(\ref{eqn.multipole2}).  The term linear in cos$\theta^*_{\eta}$ shows definite structure at all measured $Q^2$.  It was observed that as $W$ increases above where the $S_{11}$(1535) is expected to be dominant, the cos$\theta^*_{\eta}$ dependence changes dramatically.  At $W$ = 1.66 GeV it decreases monotonically with cos$\theta^*_{\eta}$, but by $W$ = 1.72 GeV the forward backward-asymmetry is reversed.  Previous experiments, at photoproduction~\cite{Renard:2000iv} and at higher $Q^2$~\cite{brasse84}, have shown the same structure in the $W$ dependence of $B$, with $B/A$ appearing to be roughly independent of $Q^2$ up to $Q^2=2.5$ GeV$^2$~\cite{Denizli:2007tq}.

\begin{figure}[!hbt]
\begin{center}
\includegraphicstwowidths{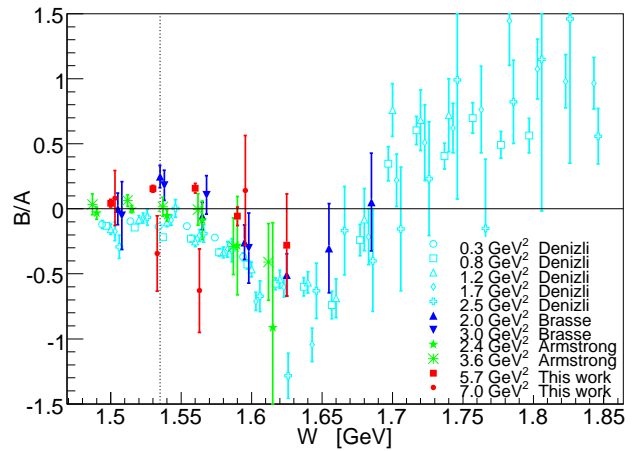}{0.48}{0.96}
\caption[Ratio of the linear cos$\theta^*_{\eta}$ term to the isotropic component]{(Colour online) The result of fits to the differential cross-section, plotted as the ratio of the linear cos$\theta^*_{\eta}$ term to the isotropic component, for the present work and other $\eta$-electroproduction data~\cite{brasse84,armstrong99,Denizli:2007tq}.  The black dotted line is drawn at $W$ = 1.535 GeV, the nominal mass for the $S_{11}$ resonance.}
\label{fig.BoA}
\end{center}
\end{figure}

The quantity $B/A$ for the present work and previously published data~\cite{Denizli:2007tq,brasse84,armstrong99} is plotted in Fig.~\ref{fig.BoA}.  Due to diminishing angular acceptance the present work does not extend above $W\sim1.65$ GeV where the ratio reaches its minimum and begins to make a rapid change from negative to positive.  For $W$ near the $S_{11}$ resonance mass (black dotted line in figure), the $B/A$ structure shows some difference between the CLAS data~\cite{Denizli:2007tq} which remains negative and data from the present work and others~\cite{brasse84,armstrong99} which do go positive, but the \textit{trend} is the same and continues to be approximately independent of $Q^2$ up to $\sim$5.8 GeV$^2$.

\begin{figure}[!hbt]
\begin{center}
\includegraphicstwowidths{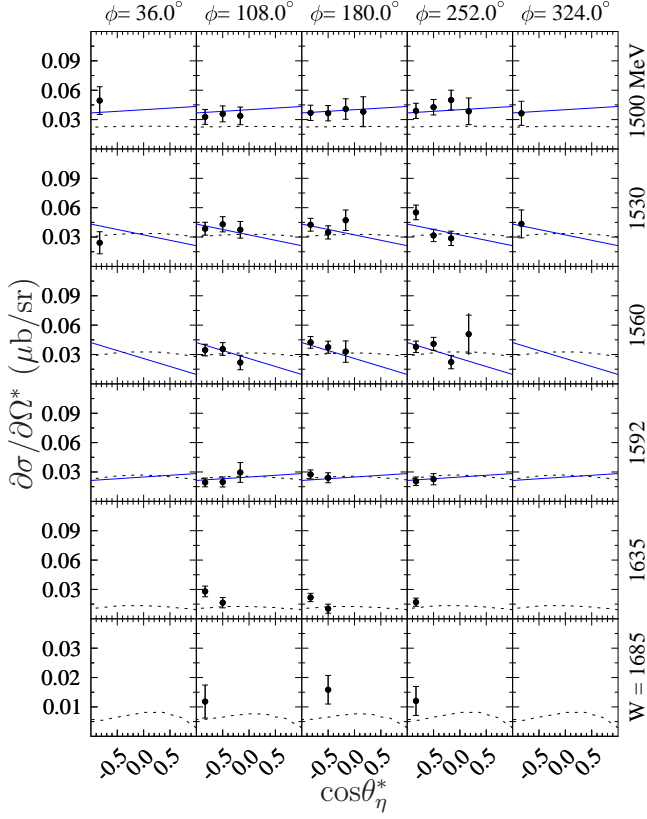}{0.48}{0.7}
\caption[Extracted $ep\rightarrow ep\eta$ differential cross-sections (higher-$Q^2$)]{(Colour online) Extracted $ep\rightarrow ep\eta$ differential cross-sections for the higher-$Q^2$ setting.  The (blue) solid curve is a fit to the data of the form ${d\sigma}/{d\Omega^*}  =  A + B~\mathrm{cos}\theta^*$.  The dashed curve is the \textsc{eta-maid} model~\cite{Chiang:2001as} at $Q^2$ = 5 GeV$^2$, projected to the appropriate $Q^2$ for each $W$ bin by the factor (5 GeV$^2/Q^2(W))^3$.  The inner error bars are statistical and the outer error bars, the quadrature sum of the statistical and systematic errors.}
\label{fig.highQ2stamps}
\end{center}
\end{figure}

The higher-$Q^2$ setting data were not amenable to the full angular fit, as can be seen in Fig.~\ref{fig.highQ2stamps}, so the fit function was truncated to ${d\sigma}/{d\Omega^*}  =  A + B~\mathrm{cos}\theta^*$ and fitted to the data.  There is large uncertainty on the extraction of $B/A$ for these data, and the results are consistent with no structure, as can be seen in Fig.~\ref{fig.BoA}.

Denizli \emph{et al.}~\cite{Denizli:2007tq} show that the rapid change in sign of $B$ could be due to a $P$ wave resonance at $W \approx 1.7$ GeV.  Specifically, a simple resonance model incorporating the $P_{11}(1710)$ could describe their data, but they do acknowledge that the $P_{13}(1720)$ is also a candidate.  The approximate $Q^2$ independence of the magnitude of this feature would imply that such a $P$ wave falls similarly slowly with $Q^2$ as the $S_{11}(1535)$.  

\begin{figure}[!hbt]
\begin{center}
\includegraphicstwowidths{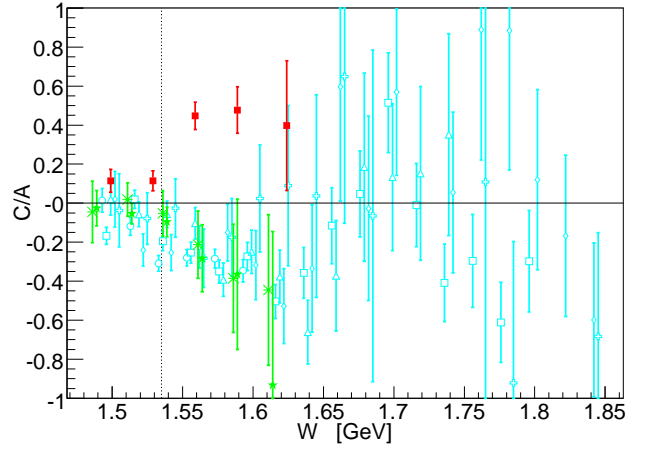}{0.48}{0.96}
\caption[Ratio of the quadratic cos$\theta^*_{\eta}$ term to the isotropic component]{(Colour online) The ratio of the quadratic cos$^2\theta^*_{\eta}$ term to the isotropic component for fits to the $\eta$-electroproduction differential cross-section, for the present work and other data~\cite{armstrong99,Denizli:2007tq}.  Symbols are the same as Fig.~\ref{fig.BoA}.}
\label{fig.CoA}
\end{center}
\end{figure}

As can be seen in Fig.~\ref{fig.fitpar},  the cos$^2\theta^*_{\eta}$ term in the angular fit to the lower-$Q^2$ data is also quite significant for $W$ above the resonance mass.  In this case, the agreement with~\cite{Denizli:2007tq} is not good, as can be seen in Fig.~\ref{fig.CoA}.  This disagreement can also clearly be seen qualitatively in Fig.~\ref{fig.lowQ2stamps} where the \textsc{eta-maid}~\cite{Chiang:2001as} curves are concave down while the new data are concave up.

\subsection{Total Cross-Section $p(e,e'p)\eta$}
\label{sec.results.total}

The total cross-section was determined from the differential cross-section in two ways.  Firstly, the total cross-section was obtained by taking the weighted mean of the differential cross-section in each $W$ bin and multiplying it by $4\pi$, where the uncertainty in the mean is the quadrature sum of the statistical and systematic errors from all the bins.  In $W$ bins where there is full coverage, this is equivalent to integrating the differential cross-section.  The total cross-section found using this method is listed in Table~\ref{tab.CS}, along with the weighted average $Q^2$ in each $W$ bin and the percentage of the 4$\pi$ c.m. angular range accepted in each $W$ bin.
\begin{table}[!hbtp]
\begin{center}
\begin{tabular}{cccc}
\hline
\hline
$\langle Q^2\rangle$  $[\frac{\mathrm{GeV}^2}{\mathrm{c}^2}$] & $W$  $[$GeV] & $\sigma$  [nb] &  $\sum\Omega^*_\eta/4\pi$ \\
\hline
5.802 &  1.50 & 831.9 $\pm$ 19.7 & 100.0\% \\
5.764 &  1.53 & 926.5 $\pm$ 20.9 & 100.0\% \\
5.704 &  1.56 & 681.5 $\pm$ 17.1 & 100.0\% \\
5.636 &  1.59 & 461.0 $\pm$ 15.0 & 93.8\% \\
5.554 &  1.62 & 336.1 $\pm$ 14.4 & 45.8\% \\
5.456 &  1.67 & 247.8 $\pm$ 14.7 & 29.2\% \\
5.353 &  1.71 & 239.0 $\pm$ 17.1 & 18.8\% \\
5.248 &  1.75 & 175.6 $\pm$ 17.8 & 16.7\% \\
5.136 &  1.78 & 160.3 $\pm$ 31.7 & 5.2\% \\
5.022 &  1.83 & 162.9 $\pm$ 27.2 & 7.3\% \\
\hline
\hline
7.064 &  1.50 & 482.1 $\pm$ 33.3 & 43.3\% \\
7.011 &  1.53 & 482.4 $\pm$ 30.5 & 36.7\% \\
6.943 &  1.56 & 437.4 $\pm$ 28.5 & 33.3\% \\
6.857 &  1.59 & 282.6 $\pm$ 25.0 & 23.3\% \\
6.746 &  1.64 & 228.9 $\pm$ 26.9 & 16.7\% \\
6.602 &  1.69 & 168.1 $\pm$ 37.5 & 10.0\% \\
6.462 &  1.74 & 230.3 $\pm$ 60.3 & 3.3\% \\
\hline
\hline
\end{tabular}
\end{center}
\caption[Total cross-section for eta electroproduction]{\label{tab.CS}Table of the total cross section, determined from the weighted average of extracted differential cross section.  The weighted average $Q^2$ and the percentage of angular coverage for each $W$ bin are also indicated.  The errors are statistical and systematic added in quadrature, and do not take into account the angular acceptance.}
\end{table}
Secondly, the fitted angular dependence, Eq.~(\ref{eqn.multipole2}) with parameters given in Table~\ref{tab.fitpar}, was integrated in each $W$ bin.  Here, the uncertainty was determined by fixing each of the six parameters to the high and low one-sigma Minuit fit values and then fitting the remaining five parameters and determining the integral.  The maximum and minimum values of the integral so determined were used to estimate the error.  This second procedure couldn't be applied to the higher-$Q^2$ setting because the sparsity of the data precluded the fitting of the full angular dependence.  The total cross sections determined in this way for each of the settings still have a $Q^2$ which varies with $W$.

The value of $\sigma_{R}$ was obtained by fitting a relativistic Breit-Wigner to the total cross-section and evaluating it at the resonance mass.  The Breit-Wigner is given by Eq.~\ref{eqn:bwQ2} and described in Section~\ref{sec.sigmodel} and the non-resonant background is modeled as $A_{\mathrm{nr}}\sqrt{W - W_{\mathrm{thr}}} + B_{\mathrm{nr}}(W - W_{\mathrm{thr}})$.  During the fit, the mean $Q^2$ value for that W bin was used.  Due to strong correlations between the parameters, especially $b_{\eta}$ and $W_R$, the branching fraction to $\eta$ was fixed at $b_{\eta}$ = 0.5  for the fits.  The uncertainty in $\sigma_R$ was estimated by individually fixing each of the Breit-Wigner parameters $W_R$ and $\Gamma_R$ to their Minuit uncertainties, redoing the fit and reevaluating $\sigma_R$.  The maximum and minimum values of $\sigma_{R}$ so determined were used to estimate the error.

\begin{figure}[!hbt]
\begin{center}
\includegraphicstwowidths{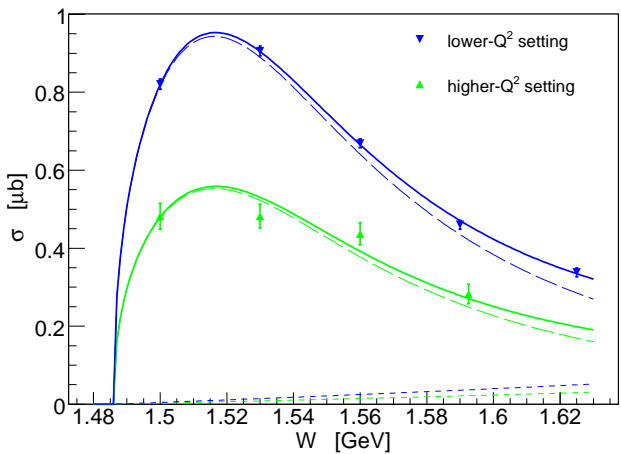}{0.48}{0.96}
\caption[Breit-Wigner fit to lower-$Q^2$ and higher-$Q^2$ data simultaneously]{(Colour online) A simultaneous fit to the lower-$Q^2$ and higher-$Q^2$ data of the sum (solid line) of a relativistic Breit-Wigner (long dash) and non-resonant background term (short dashed line).  The data are the total cross section determined from $4\pi\langle d\sigma/d\Omega^*\rangle$.  The background was constrained as described in the text.}
\label{fig.bw2waybgsQ2}
\end{center}
\end{figure}

This method worked well for the lower$Q^2$ data, with good agreement of a single Breit-Wigner to the data.  For the averaged differential cross-section a small background contribution, less than 0.5\%, was admitted under the resonance peak, while the fit to the integrated angular dependence model didn't admit any background contribution.  The higher-$Q^2$ data were amenable to such a fit since the large error bars and poor angular coverage make the parameters unreliable.  For this reason, a simultaneous fit to both settings was thus done, yielding a single set of resonance parameters.  The background was constrained to have the same $Q^2$ dependence as the data, essentially requiring it to have the same relative size.  Figure~\ref{fig.bw2waybgsQ2} shows the results of this fit, which are listed in Table~\ref{tab.ampres} along with the results of the fits to the lower-$Q^2$ data.  The shape of the fitted function is dominated by the lower-$Q^2$ data, with a background of 1.2\% at the resonance mass.  The values from this simultaneous fit are used in the further analysis.

\begin{figure}[!hbt]
\begin{center}
\includegraphicstwowidths{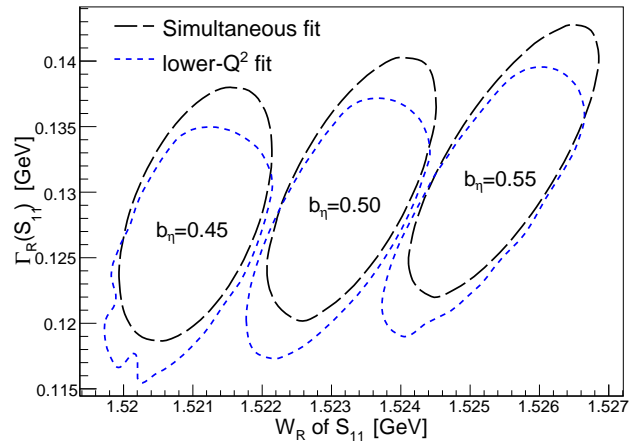}{0.48}{0.96}
\caption[1-sigma contours from Breit-Wigner fits to data]{(Colour online) Plot of the 1-sigma contours from the various Breit-Wigner fits to the data.}
\label{fig.cont}
\end{center}
\end{figure}

Both the simultaneous and the individual fits were repeated for $b_{\eta}$ = 0.45 and 0.55.  The results of these additional fits are plotted as correlation contours in Fig.~\ref{fig.cont}.  The $\sigma_{R}$ extracted from each of these additional fits was at all times well within the error quoted in Table~\ref{tab.ampres}.  It can be seen that there are correlations between $b_{\eta}$ and $W_R$ and also between $W_R$ and $\Gamma_R$.  The resonance parameters from the simultaneous fit are dominated by the lower-$Q^2$ data, as expected.

\subsection{Helicity Amplitude $A_{1/2}$ for the $S_{11}(1535)$ Resonance}
\label{sec.results.helicity}

The amplitude $A_{1/2}$ is determined from the total cross-section at the $S_{11}(1535)$ resonance mass $\sigma_R$, by Eq.~(\ref{eqn.helicity2cs}), which assumes $A_{1/2}\gg S_{1/2}$.  Using the $\sigma_R$ values obtained from the Breit-Wigner fit to the total cross-section, and those obtained in previous experiments~\cite{armstrong99,brasse84,Denizli:2007tq}, $A_{1/2}$ is determined consistently for all data with  $\Gamma_R=150$ MeV, $b_{\eta}=0.55$ and $W_R = 1535$ MeV, chosen to coincide with those used previously~\cite{armstrong99,Denizli:2007tq}.  The uncertainties in $A_{1/2}$ do not include uncertainties in $W_R$, $b_{\eta}$ or $\Gamma_R$.

\begin{figure}[!hbt]
\begin{center}
\includegraphicstwowidths{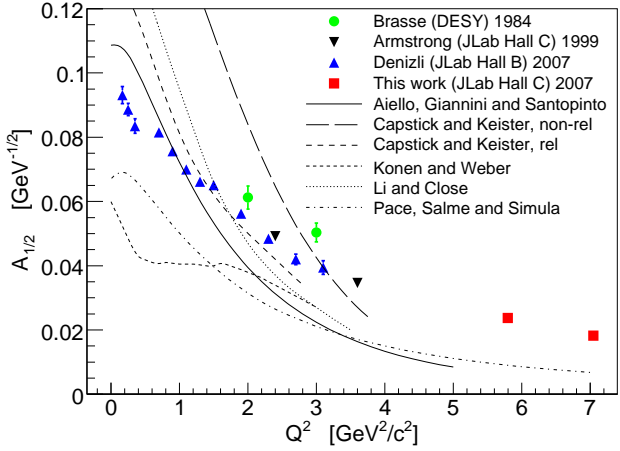}{0.48}{0.96}
\caption[The world's data for $A_p^{1/2}(S_{11})$ measured via $ep\rightarrow e'p\eta$]{(Colour online) Values for A$_{1/2}(Q^2)$ determined from $\sigma_R$ for the present and other data~\cite{armstrong99,brasse84,Denizli:2007tq} (consistently with $W_R=1.53$ GeV, $\Gamma_R=150$ MeV and $b_{\eta}=0.55$).  The curves are from Refs.~\cite{Aiello:1998xq,Capstick:1994ne,Li:1990qu,Konen:1989jp,Pace:1998pp}.}
\label{fig.helicity}
\end{center}
\end{figure}

Table~\ref{tab.ampres} summarises the parameters from the Breit-Wigner fit, the extracted total cross-section at the resonance mass, $\sigma_R$, and the extracted helicity amplitude, $A_{1/2}$.  As can be seen in Fig.~\ref{fig.helicity}, the values of A$_{1/2}$ determined in this work significantly extend the $Q^2$ range of the world's data.  The curves in the figure~\cite{Aiello:1998xq,Capstick:1994ne,Li:1990qu,Konen:1989jp,Pace:1998pp} show a huge variation in the predicted values of $A_{1/2}$.

\ifthenelse{\equal{\twocol}{false}}{
	\begin{sidewaystable}[!hbtp]
}{
	\begin{table*}[!hbtp]
}
\begin{center}
\begin{tabular}{cccccccc}
\hline
\hline
& &	& $Q^2(W_R)$ 	& $W_R$	& $\Gamma_R$	& $\sigma_R$	&  A$_{1/2}$\\
& & 	& [GeV$^2$/$c^2$] 	& [GeV]	& [GeV]	& [$\mu b$]	&  [$\times 10^{-3}$ GeV$^{5/2}$]\\
\hline	
$\int d\sigma/d\Omega^*|_{\mathrm{model}}$ & indiv. & & 5.79& $1.523\pm0.001$ & $0.125\pm0.003$ & $0.976\pm0.007$	&  $23.62\pm0.09$\\
$4\pi\langle d\sigma/d\Omega^*\rangle$ &  & 
& 5.79 & $1.523\pm0.001$ & $0.128\pm0.010$ &  $0.977\pm0.024$ & $23.63\pm0.30$ \\
\hline
$4\pi\langle d\sigma/d\Omega^*\rangle$   & simul. & Fig.~\ref{fig.bw2waybgsQ2}  & 5.79 & $1.522\pm0.001$ & $0.128\pm0.009$ & $0.943\pm0.015$ & $23.22\pm0.18$\\
 $4\pi\langle d\sigma/d\Omega^*\rangle$   &       & Fig.~\ref{fig.bw2waybgsQ2}  & 7.04 & & & $0.553\pm0.020$ & $17.79\pm0.33$\\
\hline
\hline
\end{tabular}
\end{center}
\caption[Parameters from a Breit-Wigner fit to data]{\label{tab.ampres}Parameters extracted from a relativistic Breit-Wigner fit to the data.  The values for $A_{1/2}(Q^2)$ are determined from $\sigma_R$, assuming $A_{1/2}\gg S_{1/2}$, with parameters $W_R=1.53$ GeV, $\Gamma_R=150$ MeV and $b_{\eta}=0.55$.}
\ifthenelse{\equal{\twocol}{false}}{
	\end{sidewaystable}
}{
	\end{table*}
}

\begin{figure}[!hbt]
\begin{center}
\includegraphicstwowidths{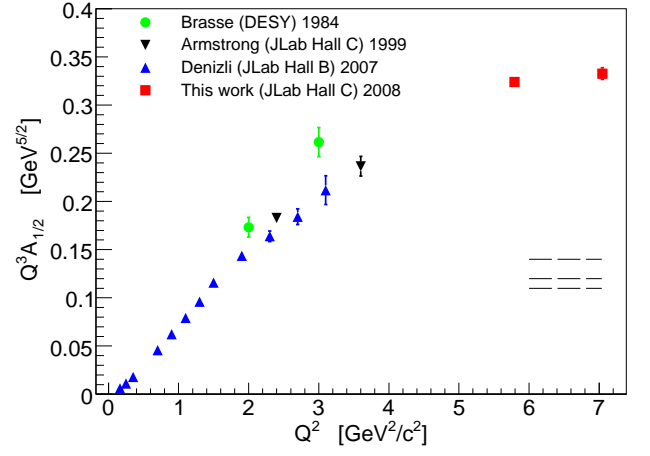}{0.48}{0.96}
\caption[The world's data for $Q^3 A_p^{1/2}$ for the $S_{11}(1535)$]
{(Colour online) The $Q^2$ dependence of $Q^3 A_{1/2}$ for $\eta$-production. Scaling in this quantity appears to begin at a photon momentum transfer of $Q^2\sim5$ GeV$^2$.  The dashed lines are a high $Q^2$, pQCD calculation from Carlson and Poor~\cite{Carlson:1988gt} using three different nucleon distribution amplitudes.}
\label{fig.scaling}
\end{center}
\end{figure}

The magnetic form-factor of the proton $G_M^p$ as published by Arnold \textit{et al.}~\cite{Arnold:1986nq} demonstrates clear scaling behaviour.  Naive dimension counting in pQCD predicts a falloff of $1/Q^4$ and the quantity of $Q^4G_M^p$ reaches a broad maximum at about $Q^2\sim$ 8 GeV$^2$ and then decreases in a gentle logarithm due to the running of the strong coupling constant $\alpha_s$.  The same arguments predict that the helicity amplitude for the $S_{11}(1535)$ decreases with $1/Q^3$.   Figure~\ref{fig.scaling} is a plot of $Q^{3}A_{1/2}$, showing that the quantity $Q^{3}A_{1/2}$ appears to begin flattening at a photon momentum transfer broadly within the range of this work, $Q^2\sim5-7$ GeV$^2$, a possible signal of the onset of pQCD scaling.  A pQCD calculation by Carlson and Poor~\cite{Carlson:1988gt}, of the magnitude of this quantity, is plotted and is a factor of $\sim3$ smaller than the data.  It has also been pointed out that such scaling may have a non-perturbative explanation~\cite{Radyushkin:1991vi,Isgur:1984jm}.

\begin{figure}[!hbt]
\begin{center}
\includegraphicstwowidths{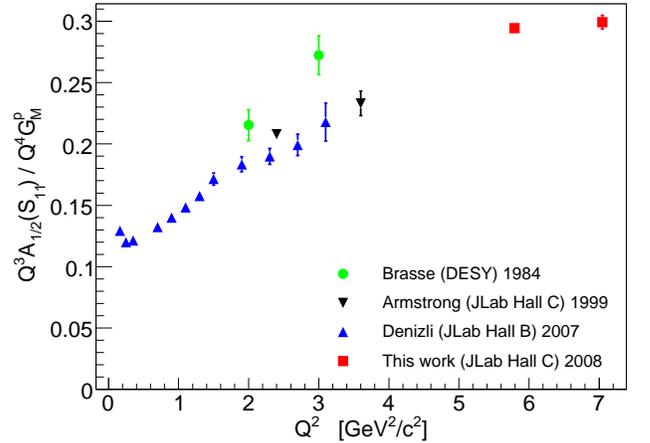}{0.48}{0.96}
\caption[The world's data for $Q^3 A_p^{1/2} / Q^4G_M^p$ for the $S_{11}(1535)$]
{(Colour online) The $Q^2$ dependence of $Q^3 A_{1/2}(S_{11})/Q^4G_M^p$ for $\eta$-production.}
\label{fig.scalingGm}
\end{center}
\end{figure}

In order to compare the behaviour of $A_{1/2}$ with the approach of $G_M^p$ to scaling, the quantity $Q^3A_{1/2}/Q^4G_M^p$ is plotted in Fig.~\ref{fig.scalingGm}.  The form of $G_M^p$ is taken from the fit by Bosted~\cite{Bosted:1994tm}.  The figure shows that the two quantities don't have the same form at low $Q^2$, and the data doesn't go high enough in $Q^2$ to know whether the two quantities begin behaving equivalently.

\section{Conclusions}
\label{sec.conc}

We have presented the results of a precise, high statistics measurement of the differential cross-section for the $e p \rightarrow e'p\eta$ exclusive process.  This is done at the highest momentum transfer to date, namely, $Q^2$ = 5.8 and 7.0 (GeV/$c$)$^2$ at the $S_{11}$ resonance mass, which is a significant extension from the previous highest at $Q^2$ = 3.6 GeV$^2$.  Data were obtained from threshold to $W = 1.8$ GeV, the $S_{11}(1535)$ dominating the channel as expected.  In the region from threshold to the $S_{11}(1535)$ resonance mass, the differential cross-section is largely isotropic---consistent with previous measurements.

The interference phenomenon in the linear cos$\theta^*_{\eta}$ term at $W$ of the $S_{11}(1535)$ resonance mass, seen in lower $Q^2$ and photoproduction data is observed here with similar strength.  The present data doesn't have sufficient angular coverage at $W\sim$ 1.7 GeV to comment meaningfully on the strong presence of a $P$ wave resonance there.  The curvature in the cos$\theta^*_{\eta}$ dependence of the differential cross-section is opposite to that of the data at lower-$Q^2$.  The helicity-conserving transition amplitude $A_{1/2}$, is extracted from the data assuming no longitudinal component ($A_{1/2}\gg S_{1/2}$).  The $Q^2$ dependence of $Q^3A_{1/2}$ seems to be flattening, consistent with the pQCD prediction, although the range of $Q^2$ is too small to verify the exact dependence.  Even if the data scale as predicted by pQCD, that is not conclusive evidence for the onset of pQCD. 

On the theoretical front, the differential cross-section will be incorporated into multi-channel, multi-resonance models, such as those by the \textsc{maid} and \textsc{ebac} groups, which should maximize the physics impact coming from these data.  Also, the inability for any one calculation to adequately describe the $Q^2$ dependence of $A_{1/2}$ leaves much to be done in understanding the structure of the $S_{11}(1535)$.  On the experimental front, more data are required to further address the questions in this paper. 

It would be nice to fill the data gap in the region between $Q^2 \sim 4$ and 5.8 GeV$^2$ to analyse the apparent change of differential cross-section shape.  Extending the data to $Q^2$ much higher than 7 GeV$^2$ will complete the study of the transition to hard-scale scattering.  Obtaining  $LT$ separated data at high $Q^2$ will enable checking of the assumption, made in this work and in the literature, that the longitudinal component is negligible.  The planned upgrade of the Jefferson Lab accelerator, to energies as high as 11 GeV, will allow exclusive $\eta$ electroproduction data to be obtained to $Q^2 \sim$ 14 GeV$^2$, and $LT$ separations at least to the $Q^2$ of this experiment.

\begin{acknowledgments}

We would like to acknowledge the support of staff and management at Jefferson Lab.

This work is supported in part by research grants from the U.S. Department of Energy (including grant DE-AC02-06CH11357), the U.S. National Science Foundation and the South African National Research Foundation.

The Southeastern Universities Research Association operates the Thomas Jefferson National Accelerator Facility under the U.S. Department of Energy contract DEAC05-84ER40150.

\end{acknowledgments}

\appendix*

\section{Tables of Differential Cross-Sections}

\linespread{1}

\begin{center}
\begin{longtable}{ccccccccc}
\caption[Extracted differential cross-section for the lower-$Q^2$ data.]{Lower-$Q^2$ extracted differential cross-section.} \label{tab.cslowQ2} \\

\hline
\hline 
$W$ & cos$\theta^*_{\eta}$ & $\phi_{\eta}$ & $\langle Q^2_{\mathrm{bin}}\rangle$ & $\langle\epsilon\rangle$ & $^\mathrm{~rad.}_\mathrm{~corr.}$ & $\frac{d^2\sigma}{d\Omega^*_{\eta}}$ & $\delta_{\mathrm{stat}}$   & $\delta_{\mathrm{syst}}$\\
$[$GeV] &                  & [deg.]        & [$\frac{\mathrm{GeV}^2}{\mathrm{c}^2}$] &      &     & [$\frac{n\mathrm{b}}{\mathrm{sr}}$]    & [$\frac{n\mathrm{b}}{\mathrm{sr}}$] & [$\frac{n\mathrm{b}}{\mathrm{sr}}$] \\
\hline 
\endfirsthead
 
$W$ & cos$\theta^*_{\eta}$ & $\phi_{\eta}$ & $\langle Q^2_{\mathrm{bin}}\rangle$ & $\langle\epsilon\rangle$ & $^\mathrm{~rad.}_\mathrm{~corr.}$ & $\frac{d^2\sigma}{d\Omega^*_{\eta}}$ & $\delta_{\mathrm{stat}}$   & $\delta_{\mathrm{syst}}$\\
$[$GeV] &                  & [deg.]        & [$\frac{\mathrm{GeV}^2}{\mathrm{c}^2}$] &      &     & [$\frac{n\mathrm{b}}{\mathrm{sr}}$]    & [$\frac{n\mathrm{b}}{\mathrm{sr}}$] & [$\frac{n\mathrm{b}}{\mathrm{sr}}$] \\
\hline 
\endhead

\endfoot

\hline
\hline
\endlastfoot
1.500  &  -0.917  &    22.5  &  5.80  &  0.427  &  1.51  &   60.6  &   8.3  &   3.6 \\ 
1.500  &  -0.917  &    67.5  &  5.80  &  0.427  &  1.52  &   72.8  &   9.2  &   6.9 \\ 
1.500  &  -0.917  &   112.5  &  5.80  &  0.426  &  1.49  &   59.1  &   8.0  &   4.6 \\ 
1.500  &  -0.917  &   157.5  &  5.80  &  0.427  &  1.52  &   72.3  &   8.9  &   5.2 \\ 
1.500  &  -0.917  &   202.5  &  5.80  &  0.426  &  1.50  &   65.1  &   8.5  &   5.5 \\ 
1.500  &  -0.917  &   247.5  &  5.80  &  0.427  &  1.52  &   72.3  &   8.8  &   6.0 \\ 
1.500  &  -0.917  &   292.5  &  5.80  &  0.426  &  1.52  &   67.0  &   8.4  &   6.2 \\ 
1.500  &  -0.917  &   337.5  &  5.80  &  0.426  &  1.49  &   77.8  &   9.4  &   7.0 \\
\end{longtable}
\end{center}

\begin{center}
\begin{longtable}{ccccccccc}
\caption[Extracted differential cross-section for the higher-$Q^2$ data.]{Higher-$Q^2$ extracted differential cross-section.} \label{tab.cshighQ2} \\

\hline
\hline 
$W$ & cos$\theta^*_{\eta}$ & $\phi_{\eta}$ & $\langle Q^2_{\mathrm{bin}}\rangle$ & $\langle\epsilon\rangle$ & $^\mathrm{~rad.}_\mathrm{~corr.}$ & $\frac{d^2\sigma}{d\Omega^*_{\eta}}$ & $\delta_{\mathrm{stat}}$   & $\delta_{\mathrm{syst}}$\\
$[$GeV] &                  & [deg.]        & [$\frac{\mathrm{GeV}^2}{\mathrm{c}^2}$] &      &     & [$\frac{n\mathrm{b}}{\mathrm{sr}}$]    & [$\frac{n\mathrm{b}}{\mathrm{sr}}$] & [$\frac{n\mathrm{b}}{\mathrm{sr}}$] \\
\hline 
\endfirsthead
 
$W$ & cos$\theta^*_{\eta}$ & $\phi_{\eta}$ & $\langle Q^2_{\mathrm{bin}}\rangle$ & $\langle\epsilon\rangle$ & $^\mathrm{~rad.}_\mathrm{~corr.}$ & $\frac{d^2\sigma}{d\Omega^*_{\eta}}$ & $\delta_{\mathrm{stat}}$   & $\delta_{\mathrm{syst}}$\\
$[$GeV] &                  & [deg.]        & [$\frac{\mathrm{GeV}^2}{\mathrm{c}^2}$] &      &     & [$\frac{n\mathrm{b}}{\mathrm{sr}}$]    & [$\frac{n\mathrm{b}}{\mathrm{sr}}$] & [$\frac{n\mathrm{b}}{\mathrm{sr}}$] \\
\hline 
\endhead
\endfoot
\hline
\hline
\endlastfoot
1.500  &  -0.833  &    36.0  &  7.04  &  0.216  &  1.52  &   49.4  &  14.2  &   3.4 \\ 
1.500  &  -0.833  &   108.0  &  7.06  &  0.211  &  1.56  &   32.7  &   7.6  &   1.5 \\ 
1.500  &  -0.833  &   180.0  &  7.07  &  0.210  &  1.57  &   36.9  &   7.8  &   2.0 \\ 
1.500  &  -0.833  &   252.0  &  7.06  &  0.210  &  1.54  &   38.9  &   7.8  &   2.3 \\ 
1.500  &  -0.833  &   324.0  &  7.04  &  0.216  &  1.53  &   36.4  &  12.3  &   2.0 \\ 
1.500  &  -0.500  &   108.0  &  7.06  &  0.211  &  1.53  &   35.8  &   8.1  &   2.0 \\ 
1.500  &  -0.500  &   180.0  &  7.07  &  0.210  &  1.53  &   36.5  &   7.8  &   1.8 \\ 
1.500  &  -0.500  &   252.0  &  7.06  &  0.211  &  1.53  &   42.6  &   7.9  &   2.1 \\
\end{longtable}
\end{center}

\bibliography{S11paper}

\end{document}